\documentclass[imslayout,preprint]{imsart} 

\usepackage[T1]{fontenc}
\usepackage{ae,aecompl}
\usepackage{graphicx}
\usepackage{amsmath,amssymb,amsthm,amsfonts}
\usepackage{enumerate}
\usepackage{color}
\usepackage{natbib}
\usepackage[unicode]{hyperref}
\usepackage[ruled,vlined]{algorithm2e}
\usepackage[table]{xcolor}

\startlocaldefs%
\def\given{{\,|\,}}
\let\hat\widehat%
\let\tilde\widetilde%
\def\Pr{{\ensuremath{\mathbb P}}}%
\def\Exp{{\ensuremath{\mathbb E}}}%
\def\logit{\ensuremath{{\rm logit}}}%
\def\argmin{\mathop{\rm arg\,min}\limits}%
\def\argmax{\mathop{\rm arg\,max}\limits}%
\def\iid{\ensuremath{\stackrel{\text{\tiny iid}}{\sim}}}%
\def\ind{\ensuremath{\stackrel{\text{\tiny ind}}{\sim}}}%
\endlocaldefs%

\begin{document}
\begin{frontmatter}

\title{Gene-proximity models for Genome-Wide Association Studies}
\runtitle{Gene-proximity models for GWAS}

\begin{aug}
\author{\fnms{Ian} \snm{Johnston}\corref{}%
\ead[label=e1]{ianj@math.bu.edu}},
\author{\fnms{Timothy} \snm{Hancock}%
\ead[label=e2]{timothy.hancock@depi.vic.gov.au}},
\author{\fnms{Hiroshi} \snm{Mamitsuka}%
\thanksref{t3}\ead[label=e3]{mami@kuicr.kyoto-u.ac.jp}},
\thankstext{t3}{Supported in part by JSPS KAKENHI grant 24300056.}
\and
\author{\fnms{Luis}  \snm{Carvalho}%
\thanksref{t4}\ead[label=e4]{lecarval@math.bu.edu}}
\thankstext{t4}{Supported in part by NSF grant DMS-1107067.}

\runauthor{I. Johnston et al.}
\affiliation{Boston University and Kyoto University}
\address{Department of Mathematics and Statistics\\ 
Boston University\\
111 Cummington Mall\\
Boston, Massachusetts, USA 02215\\
\printead{e1,e4}}
\address{
Biosciences Research Division\\
Department of Primary Industries\\
Victoria, 5 Ring Road, Bundoora 3086, Australia\\
\printead{e2}}
\address{Institute for Chemical Research\\
Kyoto University\\
Gokasho Uji-city, Kyoto, Japan 611--0011\\
\printead{e3}}
\end{aug}

\begin{abstract}
Motivated by the important problem of detecting association between genetic
markers and binary traits in genome-wide association studies, we present a
novel Bayesian model that establishes a hierarchy between markers and genes by
defining weights according to gene lengths and distances from genes to
markers. The proposed hierarchical model uses these weights to define unique
prior probabilities of association for markers based on their proximities to
genes that are believed to be relevant to the trait of interest. We use an
expectation-maximization algorithm in a filtering step to first reduce the
dimensionality of the data and then sample from the posterior distribution of
the model parameters to estimate posterior probabilities of association for
the markers. We offer practical and meaningful guidelines for the selection of
the model tuning parameters and propose a pipeline that exploits a singular
value decomposition on the raw data to make our model run efficiently on large
data sets. We demonstrate the performance of the model in simulation studies
and conclude by discussing the results of a case study using a real-world
dataset provided by the Wellcome Trust Case Control Consortium.
\end{abstract}


\end{frontmatter}

\section{Introduction}

A genome-wide association study (GWAS) aims to determine the subset of genetic
markers that is most relevant to a particular trait of interest.
From a statistical perspective, this task is usually framed as a regression
problem where the response variables are measurements of either qualitative
traits, e.g., a binary value indicating the presence or absence of
a disease, or quantitative traits, e.g., a person's blood pressure, and the
explanatory variables are the number of reference alleles present at each
marker, or single nucleotide polymorphism (SNP), as well as other covariates
of interest such as age or smoking status. Many linear models
\citep{balding2006tutorial,stephens2009bayesian} have been developed
to detect associations between SNPs and traits, but they generally suffer from
the ``large $p$, small $n$'' problem where the ratio of the number of
predictors, $p$, to the sample size, $n$, is on the order of hundreds to
thousands \citep{west03}. Moreover, other issues such as collinearity in the covariates due to linkage
disequilibrium [LD, \citep{pritchard01}], rare variants, and population
stratification result in inefficient estimation of model parameters and a loss
in statistical power to detect significant associations \citep{wang2005genome}.

A common strategy to overcome the large-$p$-small-$n$ problem in GWAS is to
forgo analyzing the SNPs jointly and to model instead them independently. Although
successful GWAS have employed this strategy \citep{burton2007genome}, multiple
hypothesis testing leads to an increase in the Type I error and the necessary
correction for this may lead to an overly conservative threshold for
statistical significance. Strategies such as grouping SNPs based on
proximities to genes \citep{wu2010powerful} or moving windows
\citep{wu2011rare} have been proposed to allow for an increase in power by
modeling SNPs jointly, but there is no universal agreement on how to define
such windows or groupings.
Similarly, other strategies include replacing a group of highly correlated
SNPs with only one of its members \citep{ioannidis2009validating}, and
removing or collapsing rare variants within a window into a score statistic
\citep{bansal2010statistical}, but again there is no agreement on how to
choose which SNPs to retain or group. Recent approaches aim at gaining more
power by pooling information across studies through meta-analysis
\citep{evangelou2013meta}.

Although significant progress has been made on GWAS since 2000, it is still a
relevant and challenging problem with goals such as modeling interactions
between SNPs, genes, and environment effects that await beyond the obstacles
already mentioned \citep{heard2010ten}. In order to move towards a unifying
framework for GWAS that allows for the large-$p$-small-$n$ problem and
the SNP-specific issues to be addressed simultaneously in a principled manner,
we propose a novel hierarchical Bayesian model that exploits spatial
relationships on the genome to define SNP-specific prior distributions on
regression parameters. More specifically, in our proposed setting we model
markers \emph{jointly}, but we explore a variable selection approach that uses
marker proximity to relevant genomic regions, such as genes, to help identify
associated SNPs. Our contributions are:

\begin{enumerate}
\item We focus on binary traits which are arguably more common to GWAS, e.g.,
case control studies, but more difficult to model due to lack of conjugacy.
To circumvent the need for a Metropolis-Hastings step when sampling from the
posterior distribution on model parameters, we use a recently proposed data
augmentation strategy for logistic regression based on latent
P\'{o}lya-Gamma random variables \citep{polson13}.

\item We perform variable selection by adopting a spike-and-slab prior
\citep{george93,ishwaran2005spike} and propose a principled way to control the
separation between the spike and slab components using a Bayesian false
discovery rate similar to \citep{whittemore2007bayesian}.

\item We use a novel weighting scheme to establish a relationship between SNPs
and genomic regions and allow for SNP-specific prior distributions on the model
parameters such that the prior probability of association for each SNP is a
function of its location on the chromosome relative to neighboring regions. 
Moreover, we allow for the ``relevance'' of a genomic region to contribute to
the effect it has on its neighboring SNPs and consider ``relevance'' values
calculated based on previous GWAS results in the literature, e.g.
see~\citep{MalaCards}.

\item Before sampling from the posterior space using Gibbs sampling, we use an
expectation-maximization [EM, \citep{dempster77}] algorithm in a filtering step
to reduce the number of candidate markers in a manner akin to distilled
sensing~\citep{haupt2011distilled}. By investigating the update equations for
the EM algorithm, we suggest meaningful values to tune the hyperprior
parameters of our model and illustrate the induced relationship between SNPs
and genomic regions.

\item We derive a more flexible centroid estimator~\citep{carvalho08} for SNP
associations that is parameterized by a sensitivity-specificity trade-off. We
discuss the relation between this parameter and the prior specification when
obtaining estimates of model parameters.
\end{enumerate}

We start by describing previous work and stating our contributions in
Section~\ref{sec:previous}. In Section~\ref{sec:model} we define our Spatial
Boost model and the novel relationship between SNPs and genes.
In Section~\ref{sec:fit} we describe how to fit the model using a combination
of a filtering step that exploits an EM filtering step and Gibbs sampling. We
provide guidelines for the selection of model tuning parameters in
Section~\ref{sec:guidelines}. We then illustrate the performance of the model
on simulated data using real SNPs in Section~\ref{sec:study} and apply the
model to a real-world GWAS data set provided by the Wellcome Trust Case
Control Consortium (WTCCC) in Section~\ref{sec:case}. Finally, we conclude
with a discussion on future extensions to this work in
Section~\ref{sec:conclusion}.

\section{Previous and Related Work}
\label{sec:previous}
A common solution to large-$p$-small-$n$ problems is to use penalized
regression models such as ridge regression \citep{hoerl70}, LASSO
\citep{tibshirani96}, or elastic net \citep{zou05}. These solutions can be
shown to be equivalent, from a Bayesian perspective, to maximum \emph{a
posteriori} (MAP) estimators under appropriate prior specifications.
For instance, for LASSO, the $L_1$ penalty can be translated into a Laplace
prior. However, since LASSO produces biased estimates of the model parameters
and tends to select only one parameter in a group of correlated parameters
\citep{zou05}, it is not suitable for GWAS.\@

Techniques like group LASSO, fused LASSO \citep{tibshirani05}, or sparse group
LASSO \citep{friedman10} further attempt to account for the structure of genes
and markers or linkage disequilibrium by assigning SNPs to groups based on
criteria such as gene membership and then placing additional penalties on the
$L_1$ norm of the vector of coefficients for each group, or on the $L_1$ norm
of the difference in coefficients of consecutive SNPs. However, it is
difficult to define gene membership universally since genes have varying
lengths and may overlap with each other; moreover, the penalty on the $L_1$
norm of the difference in consecutive SNPs neglects any information contained
in the genomic distance between them. 

It may be possible to develop additional, effective penalty terms within
models, such as $L_1$ and $L_2$, to address the issues present in GWAS data in
a penalized regression framework, but because genotypes are more correlated
for markers that are close in genomic space due to linkage disequilibrium, the
most effective penalties would need to capture the relevance of a particular
SNP as a function of its location on the genome. Moreover, since it is
typically easier to study the biological function of genes, we are
particularly interested in SNPs that lie close to genes
\citep{jorgenson2006gene}; as a result, the most desirable penalties would
likely be SNP-specific.
We accomplish this by exploiting biological knowledge about the structure of
the genome to set SNP-specific prior distributions on the model parameters in
a hierarchical Bayesian model.
%Since the fundamental principle of Bayesian statistics is to incorporate prior
%knowledge when fitting model parameters, it is arguably more natural to
%exploit external knowledge from biology about the structure of the genome
%through prior distributions on the model parameters in a Bayesian framework.

Researchers have considered hierarchical Bayesian models for variable
selection in GWAS and other large scale problems (e.g.
\citep{guan2011bayesian,zhou2013polygenic}). Some recent models exploit
Bayesian methods in particular to allow for data-driven SNP-specific prior
distributions \citep{habier11} which depend on a random variable that
describes the proportion of SNPs to be selected. These approaches have adopted
a continuous spike-and-slab prior distribution
\citep{george93,ishwaran2005spike} on the model parameters, set an
inverse-gamma prior distribution on the variance of the spike component of the
prior, and control the difference in the variance of the spike and slab
components of the prior using a tuning parameter.

To incorporate external information in a hierarchical Bayesian model,
researchers analyzing a different kind of data, gene expression levels, have
recently considered relating a linear combination of a set of predictor-level
covariates that quantify the relationships between the genes to their prior
probabilities of association through a probit link function
\citep{peng2013integrative}. This formulation leads to a second-stage probit
regression on the probability that any gene is associated with a trait of
interest using a set of predictor-level covariates that could be, for
instance, indicator variables of molecular pathway membership. In our model,
we propose a special case of this formulation tailored for GWAS
data where: (i) we use the logit link instead of the probit link, (ii) the
predictor-level covariates are spatial weights that quantify a SNP's position
on the genome relative to neighboring genes, and (iii) the coefficients of
each of the predictor-level covariates are numerical scores that quantify the
relevance of a particular gene to the trait of interest. 

Fitting a penalized model to a large data set (e.g. $p \ge 100,\!000$) is
computationally intense and thus so is the process of selecting an optimal
value for any tuning parameters. Side-stepping this problem, some
researchers have had success in applying a suite of penalty terms (e.g. LASSO,
Adaptive LASSO, NEG, MCP, LOG) to a \emph{pre-screened} subset of markers
\citep{hoffman2013puma} and investigating the concordance of significant
markers across each of the final models. Although a pre-screening of markers
from a marginal regression would ideally retain almost all of the relevant
variables, penalized models such as LASSO could likely be improved by
using a larger number of SNPs than those which pass an initial screening
step (e.g.\ a genome-wide significance threshold) \citep{kooperberg2010risk}.

\section{Spatial Boost Model}
\label{sec:model}
We perform Bayesian variable selection by analyzing binary traits and using
the structure of the genome to dynamically define the prior probabilities of
association for the SNPs. Our data are the binary responses $\mathbf{y} \in
{\{0,1\}}^n$ for $n$ individuals and genotypes
${\mathbf{x}}_i \in {\{0,1,2\}}^p$
for $p$ markers per individual, where $x_{ij}$ codes the number of
\emph{minor} alleles in the $i$-th individual for the $j$-th marker.
For the likelihood of the data, we consider the logistic regression: 
\begin{equation}
\label{eq:lhood}
y_i \given \mathbf{x}_i, \beta \ind
\text{\sf Bern}\big(\logit^{-1}(\beta_0 + \mathbf{x}_i^\top \beta)\big),
\quad \text{for~} i = 1, \ldots, n.
\end{equation}
We note that GWA studies are usually \emph{retrospective}, i.e.\ cases and
controls are selected irrespectively of their history or genotypes; however,
as \citet{mccullagh1989generalized} point out, coefficient estimates for
$\beta$ are not affected by the sampling design under a logistic regression.
Thus, from now on, to alleviate the notation we extend $\mathbf{x}_i$ to
incorporate the intercept, $\mathbf{x}_i = (x_{i0} = 1, x_{i1}, \ldots,
x_{ip})$, and also set $\beta = (\beta_0, \beta_1, \ldots, \beta_p)$.

We use latent variables $\theta \in {\{0,1\}}^p$ and a continuous
spike-and-slab prior distribution for the model parameters with the positive
constant $\kappa > 1$ denoting the separation between the variance of the
spike and the slab components:
\begin{equation}
\label{eq:betaprior}
\beta_j \given \theta_j, \sigma^2 \ind \text{\sf N}\big(0,
\sigma^2 [\theta_j \kappa + (1 - \theta_j) ]\big),
\quad \text{for~} j = 1, \ldots, p.
\end{equation}
For the intercept, we set $\beta_0 \sim N(0, \sigma^2 \kappa)$ or,
equivalently, we define $\theta_0 = 1$ and include $j = 0$
in~\eqref{eq:betaprior}.
In the original spike-and-slab prior distribution, the slab component is a
normal distribution centered at zero with a large variance or even a
diffuse uniform distribution and the spike component is a degenerate
distribution at zero~\citep{mitchell1988bayesian}.
This setup results in exact variable selection through the use of the
$\theta_j$'s, since $\theta_j = 0$ would imply that the $j$-th SNP's
coefficient is exactly equal to zero. Here we use the continuous version of
the spike-and-slab distribution~\citep{george93,ishwaran2005spike} that became
more popular by avoiding the spike discontinuities at zero and thus allowing
for a relaxed form of variable selection that lends itself easily to an EM
algorithm (see Section~\ref{sec:em}).

For the variance $\sigma^2$ of the spike component in~\eqref{eq:betaprior} we
adopt an inverse Gamma prior distribution,
$\sigma^2 \sim \text{\sf IG}(\nu, \lambda)$.
We expect $\sigma^2$ to be reasonably small with high probability in order to
enforce the desired regularization that distinguishes selected markers from
non-associated markers. Thus, we recommend choosing $\nu$ and $\lambda$ so
that the prior expected value of $\sigma^2$ is small.

In the prior distribution for $\theta_j$, we incorporate information from
relevant genomic regions. The most common instance of such regions are
\emph{genes}, and so we focus on these regions in what follows. Thus, given a
list of $G$ genes with gene \emph{relevances} (see Section~\ref{sec:generel}
for some choices of definitions), $\mathbf{r} = [r_1, r_2, \ldots, r_G]$, and
weights, $\mathbf{w}_j(\phi) = [w_{j,1}, w_{j,2}, \ldots, w_{j,G}]$, the prior
on $\theta_j$ is
\begin{equation}
\label{eq:thetaprior}
\theta_j \ind \text{\sf Bern}\big(
\logit^{-1}(\xi_0 + \xi_1 {\mathbf{w}_j(\phi)}^\top\mathbf{r})
\big),\quad \text{for~} j = 1, \ldots, p.
\end{equation}
The weights $\mathbf{w}_j$ are defined using the structure of the SNPs and
genes and aim to account for gene lengths and their proximity to markers as a
function of a spatial parameter $\phi$, as we see in more detail next.

\begin{figure}
\includegraphics[width=.95\textwidth]{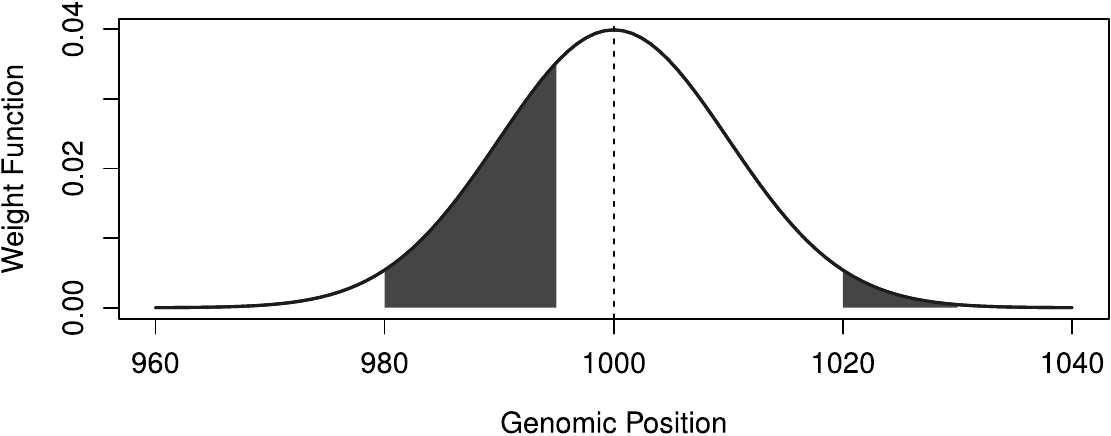}
\caption{Gene weight example: for the $j$-th SNP at position $s_j = 1,\!000$
and two surrounding genes $a$ and $b$ spanning $(980, 995)$ and $(1020, 1030)$
we obtain, if setting $\phi = 10$, weights of $w_{j,a} = 0.29$ and $w_{j,b} =
0.02$, respectively.}
\label{fig:geneweight}
\end{figure}

\subsection{Gene Weights}
\label{sec:geneweights}
To control how much a gene can contribute to the prior probability of
association for a SNP based on the gene's length and distance to that SNP we
introduce a \emph{range} parameter $\phi > 0$. Consider a gene $g$ that spans
genomic positions $g_l$ to $g_r$, and the $j$-th marker at genomic position
$s_j$; the gene weight $w_{j,g}$ is then
\[
w_{j,g} = \int_{g_l}^{g_r} \dfrac{1}{\sqrt{2\pi \phi^2}} 
\text{exp}\Bigg\{-\dfrac{{(x - s_j)}^2}{2 \phi^2} \Bigg\}\,\mathrm{d}x.
\]
Generating gene weights for a particular SNP is equivalent to centering a
Gaussian curve at that SNP's position on the genome with standard deviation
equal to $\phi$ and computing the area under that curve between the start and
end points of each gene. Figure~\ref{fig:geneweight} shows an example.
As $\phi \to 0$, the weight that each gene contributes to a particular SNP
becomes an indicator function for whether or not it covers that SNP;\@ as
$\phi \to \infty$, the weights decay to zero. Intermediate values of $\phi$
allow then for a variety of weights in $[0, 1]$ that encode \emph{spatial}
information about gene lengths and gene proximities to SNPs. In
Section~\ref{sec:phi} we discuss a method to select $\phi$.

According to~\eqref{eq:thetaprior}, it might be possible for multiple,
possibly overlapping, genes that are proximal to SNP $j$ to boost $\theta_j$.
To avoid this effect, we take two precautions. First, we break genes into
non-overlapping genomic blocks and define the relevance of a block as the mean
gene relevance of all genes that cover the block.
Second, we normalize the gene weight contributions to $\theta_j$
in~\eqref{eq:thetaprior}, ${\mathbf{w}_j(\phi)}^\top \mathbf{r}$, such that
$\max_j {\mathbf{w}_j(\phi)}^\top \mathbf{r} = 1$. This way, it is possible to
compare estimates of $\xi_1$ across different gene weight and relevance
schemes.

\subsection{Gene Relevances}
\label{sec:generel}
We allow for the further strengthening or diminishing of particular gene
weights using gene relevances $\mathbf{r}$. If we set $\mathbf{r} =
\mathbf{1}_G$ and allow for all genes to be uniformly relevant, then we have
a ``non-informative'' case. Alternatively, if we have some reason to believe
that certain genes are more relevant to a particular trait than others, for
instance on the basis of previous research or prior knowledge from an expert,
then we can encode these beliefs through $\mathbf{r}$. In particular, we
recommend using either text-mining techniques, e.g. \citep{al2010text}, to
quantify the relevance of a gene to a particular disease based on citation
counts in the literature, or relevance scores compiled from search hits and
citation linking the trait of interest to genes, e.g. \citep{MalaCards}.

\section{Model Fitting and Inference}
\label{sec:fit}
The ultimate goal of our model is to perform inference on the posterior
probability of association for SNPs. However, these probabilities are not
available in closed form, and so we must resort to Markov chain Monte Carlo
techniques such as Gibbs sampling to draw samples from the posterior
distributions of the model parameters and use them to estimate
$\Pr(\theta_j = 1 \given \mathbf{y})$. Unfortunately, these techniques can be
slow to iterate and converge, especially when the number of model parameters
is large \citep{cowles1996markov}. Thus, to make our model more
computationally feasible, we propose first filtering out markers to reduce the
size of the original dataset in a strategy similar to distilled sensing
\citep{haupt2011distilled}, and then applying a Gibbs sampler to only the
remaining SNPs.

To this end, we design an EM algorithm based on the hierarchical model above
that uses all SNP data simultaneously to quickly find an approximate mode of
the posterior distribution on $\beta$ and $\sigma^2$ while regarding $\theta$
as missing data. Then, for the filtering step, we iterate between (1) removing a
fraction of the markers that have the lowest conditional probabilities of
association and (2) refitting using the EM procedure until the predictions of the
filtered model degrade. In our analyses we filtered $25\%$ of the markers at
each iteration to arrive at estimates $\beta^*$ and
stopped if $\max_i |y_i - \logit^{-1}(\mathbf{x}_i^\top \beta^*)| > 0.5$.
Next, we discuss the EM algorithm and the Gibbs sampler, and offer
guidelines for selecting the other parameters of the model in
Section~\ref{sec:guidelines}.

\subsection{EM algorithm}
\label{sec:em}
We treat $\theta$ as a latent parameter and build an EM algorithm
accordingly.
If $\ell(\mathbf{y}, \theta, \beta, \sigma^2) = \log\Pr(\mathbf{y}, \theta,
\beta, \sigma^2)$ then for the M-steps on $\beta$ and $\sigma^2$ we maximize
the expected log joint
$Q(\beta, \sigma^2; \beta^{(t)}, {(\sigma^2)}^{(t)})
= \Exp_{\theta \given \mathbf{y}, X; \beta^{(t)}, {(\sigma^2)}^{(t)}}
[\ell(\mathbf{y}, \theta, \beta, \sigma^2)]$. The log joint distribution
$\ell$, up to a normalizing constant, is
\begin{multline}
\label{eq:l}
\ell(\mathbf{y},\theta,\beta,\sigma^2) = \sum_{i = 1}^n
y_i \mathbf{x}_i^\top \beta - \log(1 + \exp\{\mathbf{x}_i^\top \beta\}) \\
-\dfrac{p+1}{2} \log\sigma^2 - \dfrac{1}{2\sigma^2}
\sum_{j=0}^p \beta_j^2 \Bigg(\dfrac{\theta_j}{\kappa} + 1 - \theta_j\Bigg)
-(\nu + 1)\log\sigma^2 - \dfrac{\lambda}{\sigma^2},
\end{multline}
and so, at the $t$-th iteration of the procedure, for the E-step we just need
$\langle \theta_j \rangle^{(t)} \doteq
\Exp_{\theta \given \mathbf{y}; \beta^{(t)}, {(\sigma^2)}^{(t)}}[\theta_j]$.
But since
\[
\langle \theta_j \rangle =
\Pr(\theta_j = 1 \given \mathbf{y}, \beta, \sigma^2) =
\dfrac{\Pr(\theta_j = 1, \beta_j \given \sigma^2)}
{\Pr(\theta_j = 0, \beta_j \given \sigma^2) +
\Pr(\theta_j = 1, \beta_j \given \sigma^2)},
\]
then
\begin{equation}
\label{eq:etheta}
\logit\langle \theta_j \rangle =
\log \dfrac{\Pr(\theta_j = 1, \beta_j \given \sigma^2)}%
{\Pr(\theta_j = 0, \beta_j \given \sigma^2)} = 
-\dfrac{1}{2}\log\kappa - \dfrac{\beta_j^2}{2\sigma^2}
\Bigg(\dfrac{1}{\kappa} - 1\Bigg)
+ \xi_0 + \xi_1 \mathbf{w}_j^\top \mathbf{r}
\end{equation}
for $j = 1, \ldots, p$ and $\langle\theta_0\rangle \doteq 1$.

To update $\beta$ and $\sigma^2$ we employ conditional maximization steps
\citep{mengrubin93}, similar to cyclic gradient descent. From~\eqref{eq:l} 
we see that the update for $\sigma^2$ follows immediately from the mode of an
inverse gamma distribution conditional on $\beta^{(t)}$:
\begin{equation}
\label{eq:msigma}
{(\sigma^2)}^{(t+1)} =
\frac{\dfrac{1}{2}\displaystyle\sum_{j = 0}^p
{(\beta_j^{(t)})}^2
\Bigg(\dfrac{\langle\theta_j\rangle^{(t)}}{\kappa}
  + 1 - \langle\theta_j\rangle^{(t)}\Bigg) + \lambda}
{\dfrac{p + 1}{2} + \nu + 1}.
\end{equation}

The terms in~\eqref{eq:l} that depend on $\beta$ come from the log likelihood
of $\mathbf{y}$ and from the expected prior on $\beta$, $\beta \sim N(0,
\Sigma^{(t)})$, where
\[
\Sigma^{(t)} = \text{Diag}\Bigg(\dfrac{\sigma^2}
{\langle\theta_j\rangle^{(t)}/\kappa + 1 - \langle\theta_j\rangle^{(t)}}
\Bigg).
\]
Since updating $\beta$ is equivalent here to fitting a ridge regularized
logistic regression, we exploit the usual iteratively reweighted least squares
(IRLS) algorithm \citep{maccullagh1989generalized}.
Setting $\mu^{(t)}$ as the vector of expected responses with
$\mu^{(t)}_i = \logit^{-1}(\mathbf{x}_i^\top \beta^{(t)})$ and
$W^{(t)} = \text{Diag}(\mu^{(t)}_i(1 - \mu^{(t)}_i))$ as
the variance weights, the update for $\beta$ is then
\begin{equation}
\label{eq:mbeta}
\beta^{(t+1)} = {(X^\top W^{(t)}X + {(\Sigma^{(t)})}^{-1})}^{-1}
\big(X^\top W^{(t)}X\beta^{(t)} + X^\top (\mathbf{y} - \mu^{(t)})\big),
\end{equation}
where we substitute ${(\sigma^2)}^{(t)}$ for $\sigma^2$ in the definition of
$\Sigma^{(t)}$.

\subsubsection*{Rank truncation of design matrix}
Computing and storing the inverse of the $(p+1)$-by-$(p+1)$ matrix
$X^\top W^{(t)}X + {(\Sigma^{(t)})}^{-1}$ in~\eqref{eq:mbeta} is expensive
since $p$ is large. To alleviate this problem, we replace $X$ with a rank
truncated version based on its singular value decomposition $X = UDV^\top$.
More specifically, we take the top $l$ singular values and their respective
left and right singular vectors, and so, if $D = \text{Diag}(d_i)$ and
$\mathbf{u}_i$ and $\mathbf{v}_i$ are the $i$-th left and right singular
vectors respectively,
\[
X = UDV^\top
= \sum_{i=1}^n d_i \mathbf{u}_i \mathbf{v}_i^\top
\approx \sum_{i = 1}^l d_i \mathbf{u}_i \mathbf{v}_i^\top
= U_{(l)} D_{(l)} V_{(l)}^\top,
\]
where $D_{(l)}$ is the $l$-th order diagonal matrix with the top $l$ singular
values and $U_{(l)}$ ($n$-by-$l$) and $V_{(l)}$ ($(p+1)$-by-$l$) contain the
respective left and right singular vectors. We select $l$ by controlling the
mean squared error: $l$ should be large enough such that
$\|X - U_{(l)} D_{(l)} V_{(l)}^\top\|_F / (n(p+1)) < 0.01$.

Since
$X^\top W^{(t)}X \approx
V_{(l)} D_{(l)} U_{(l)}^\top W^{(t)} U_{(l)} D_{(l)} V_{(l)}^\top$,
we profit from the rank truncation by defining the (upper) Cholesky
factor $C_w$ of $D_{(l)} U_{(l)}^\top W^{(t)} U_{(l)} D_{(l)}$ and
$S = C_w V_{(l)}^\top$ so that
\begin{equation}
\label{eq:invX}
\begin{split}
{(X^\top W^{(t)}X + {(\Sigma^{(t)})}^{-1})}^{-1} &\approx
{(S^\top S + {(\Sigma^{(t)})}^{-1})}^{-1} \\
&= \Sigma^{(t)} -
\Sigma^{(t)}S^\top {(I_l + S\Sigma^{(t)}S^\top)}^{-1}S\Sigma^{(t)}
\end{split}
\end{equation}
by the Kailath variant of the Woodbury identity \citep{petersen08}.
Now we just need to store and compute the inverse of the $l$-th order square
matrix $I_l + S\Sigma^{(t)}S^\top$ to obtain the updated $\beta^{(t+1)}$
in~\eqref{eq:mbeta}.

\subsection{Gibbs sampler}
After obtaining results from the EM filtering procedure, we proceed to analyze
the filtered dataset by sampling from the joint posterior
$\Pr(\theta, \beta, \sigma^2 \given \mathbf{y})$ using Gibbs sampling. We
iterate sampling from the conditional distributions
\[
[\sigma^2 \given \theta, \beta, \mathbf{y}], \quad
[\theta \given \beta, \sigma^2, \mathbf{y}], \quad \text{and} \quad
[\beta \given \theta, \sigma^2, \mathbf{y}]
\]
until assessed convergence.

We start by taking advantage of the conjugate prior for $\sigma^2$ and draw
each new sample from 
\[
\sigma^2 \given \theta, \beta, \mathbf{y} \sim
\text{\sf IG}\Bigg(\nu + \dfrac{p+1}{2},\,
\lambda + \dfrac{1}{2}\sum_{j=0}^p
\beta_j^2 \Big(\dfrac{\theta_j}{\kappa} + 1 - \theta_j\Big)\Bigg).
\]
Sampling $\theta$ is also straightforward: since the $\theta_j$ are
independent given $\beta_j$,
\[
\theta_j \given \beta, \sigma^2, \mathbf{y} \ind
\text{\sf Bern}(\langle\theta_j\rangle), \quad \text{for~}j = 1, \ldots, p,
\]
with $\langle\theta_j\rangle$ as in~\eqref{eq:etheta}. Sampling $\beta$,
however, is more challenging since there is no closed-form distribution based
on a logistic regression, but we use a data augmentation scheme proposed by
\citet{polson13}. This method has been noted to perform well when the model
has a complex prior structure and the data have a group structure and so we
believe it is appropriate for the Spatial Boost model. 

Thus, to sample $\beta$ conditional on $\theta$, $\sigma^2$, and $\mathbf{y}$
we first sample latent variables $\omega$ from a P\'{o}lya-Gamma distribution,
\[
\omega_i \given \beta \sim \text{\sf PG}(1, \mathbf{x}_i^\top\beta),
\quad i=1,\ldots,n,
\]
and then, setting $\Omega = \text{Diag}(\omega_i)$,
$\Sigma = \text{Diag}(\sigma^2(\theta_j \kappa + 1 - \theta_j))$,
and $V_{\beta} = X^\top \Omega X + \Sigma^{-1}$, sample
\[
\beta \given \omega, \theta, \sigma^2, \mathbf{y} \sim
N(V_{\beta}^{-1} X^\top(\mathbf{y} - 0.5\cdot \mathbf{1}_n),\,
V_{\beta}^{-1}).
\]
We note that the same rank truncation used in the EM algorithm from the
previous section works here, and we gain more computational efficiency by
using an identity similar to~\eqref{eq:invX} when computing and storing
$V_{\beta}^{-1}$.

\subsection{Centroid estimation}
To conduct inference on $\theta$ we follow statistical decision theory
\citep{berger85} and define an estimator based on a generalized Hamming loss
function $H(\theta, \tilde{\theta}) = \sum_{j=1}^p h(\theta_j,
\tilde{\theta}_j)$,
\begin{equation}
\label{eq:centroid0}
\hat{\theta}_C = \argmin_{\tilde{\theta} \in {\{0,1\}}^p}
\Exp_{\theta \given \mathbf{y}} \big[ H(\theta, \tilde{\theta}) \big]
= \argmin_{\tilde{\theta} \in {\{0,1\}}^p}
\Exp_{\theta \given \mathbf{y}} \Bigg[
\sum_{j=1}^p h(\theta_j, \tilde{\theta}_j) \Bigg].
\end{equation}
We assume that $h$ has symmetric error penalties, $h(0,1) = h(1,0)$ and that
$h(1,0) > \max\{h(0,0), h(1,1)\}$, that is, the loss for a false positive or
negative is higher than for a true positive and true negative.
In this case, we can define a \emph{gain} function $g$ by subtracting each
entry in $h$ from $h(1,0)$ and dividing by $h(1,0) - h(0,0)$:
\[
g(\theta_j, \tilde{\theta}_j) = \begin{cases}
1,& \theta_j = \tilde{\theta}_j = 0, \\
0,& \theta_j \ne \tilde{\theta}_j, \\
\gamma \doteq \dfrac{h(1,0) - h(1,1)}{h(1,0) - h(0,0)},&
\theta_j = \tilde{\theta}_j = 1. \\
\end{cases}
\]
Gain $\gamma > 0$ represents a sensitivity-specificity trade-off; if $h(0,0) =
h(1,1)$, that is, if true positives and negatives have the same relevance,
then $\gamma = 1$.

Let us define the marginal posteriors
$\pi_j \doteq \Pr(\theta_j = 1 \given \mathbf{y})$.
The above estimator is then equivalent to
\begin{multline*}
\hat{\theta}_C = \argmax_{\tilde{\theta} \in {\{0,1\}}^p}
\Exp_{\theta \given \mathbf{y}} \Bigg[
\sum_{j=1}^p g(\theta_j, \tilde{\theta}_j) \Bigg] \\
= \argmax_{\tilde{\theta} \in {\{0,1\}}^p}
\sum_{j=1}^p (1 - \tilde{\theta}_j)(1 - \pi_j)
+ \gamma \tilde{\theta}_j \theta_j 
= \argmax_{\tilde{\theta} \in {\{0,1\}}^p}
\sum_{j=1}^p \Bigg(\pi_j - \dfrac{1}{1 + \gamma}\Bigg)
\tilde{\theta}_j,
\end{multline*}
which can be obtained position-wise,
\begin{equation}
\label{eq:centroid}
{(\hat{\theta}_C)}_j = I\Bigg(\pi_j - \dfrac{1}{1 + \gamma} \geq 0 \Bigg).
\end{equation}

The estimator in~\eqref{eq:centroid0} is known as the \emph{centroid
estimator}; in contrast to maximum \emph{a posteriori} (MAP) estimators that
simply identify the highest peak in a posterior distribution, centroid
estimators can be shown to be closer to the mean than to a mode of the
posterior space, and so offer a better summary of the posterior distribution
\citep{carvalho08}. Related formulations of centroid estimation for binary
spaces in~\eqref{eq:centroid} have been proposed in many bioinformatics
applications in the context of maximum expected accuracy \citep{hamada12}.
Moreover, if $\gamma = 1$ then $\hat{\theta}_C$ is simply a consensus
estimator and coincides with the median probability model estimator of
\citet{barbieri04}.

Finally, we note that the centroid estimator can be readily obtained from MCMC
samples $\theta^{(1)}, \ldots, \theta^{(N)}$ since we just need to estimate
the marginal posterior probabilities
$\hat{\pi}_j = \sum_{s=1}^N \theta^{(s)}_j / N$ and apply them
to~\eqref{eq:centroid}.

\section{Guidelines for Selecting Prior Parameters}
\label{sec:guidelines}
Since genome-wide association is a large-$p$-small-$n$ problem, we rely
on adequate priors to guide the inference and overcome ill-posedness.
In this section we provide guidelines for selecting hyperpriors $\kappa$ in
the slab variance of $\beta$, and $\phi$, $\xi_0$, and $\xi_1$ in the prior
for $\theta$.

\subsection{Selecting $\phi$}
\label{sec:phi}
Biologically, some locations within a chromosome may be more prone to
recombination events and consequently to relatively higher
linkage disequilibrium. LD can be characterized as correlation in the
genotypes, and since we analyze the entire genome, high correlation in markers
within a chromosome often results in poor coefficient estimates for the
logistic regression model in~\ref{eq:lhood}. To account for potentially
varying spatial relationships across the genome, we exploit the typical
correlation pattern in GWAS data sets to suggest a value for $\phi$ that
properly encodes the spatial relationship between markers and genes in a
particular region as a function of genomic distance. To this end, we propose
the following procedure to select $\phi$:

\begin{enumerate}
\item Divide each chromosome into regions such that the distance between the
SNPs in adjacent regions is at least the average length of a human gene, or
$30,\!000$ base pairs \citep{technology2002handy}. The resulting regions will
be, on average, at least a gene's distance apart from each other and may
possibly exhibit different patterns of correlation.

\item Merge together any adjacent regions that cover the same gene. Although
the value of $\phi$ depends on each region, we want the meaning of the weights
assigned from a particular gene to SNPs in the Spatial Boost model to be
consistent across regions. As a practical example, by applying the first two
steps of the pre-processing procedure on chromosome~1, we obtain 1,299 windows
of varying sizes ranging from 1 to 300 markers.

\item Iterate over each region and select a value of $\phi$ that best fits the
magnitude of the genotype correlation between any given pair of SNPs as a
function of the distance between them. We propose using the normal curve
given in the definition of the gene weights to first fit the magnitudes, and
then using the mean squared error between the magnitudes in the sample
correlation matrix of a region and the magnitudes in the fitted correlation
matrix as a metric to decide the optimal value of $\phi$. In particular,
given two SNPs located at positions $s_i$ and $s_j$, we relate the magnitude
of the correlation between SNPs $i$ and $j$ to the area
\[
|\rho_{i,j}|(\phi) = 2 \Phi\Bigg( -\dfrac{|s_i - s_j|}{\phi} \Bigg),
\]
where $\Phi$ is the standard normal cumulative function.

\begin{figure}
\includegraphics[width=.95\textwidth]{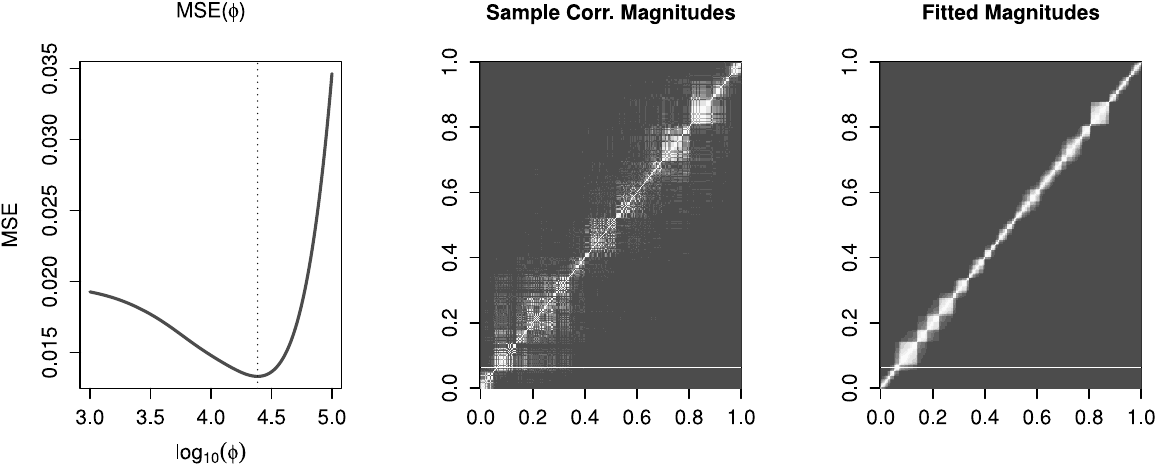}
\caption{Example of selection of $\phi$: when using the proposed values of
$|\rho_{i,j}|$ to fit the sample correlation magnitudes, we obtain an optimal
choice of $\phi = 13,\!530$ for a random window.}
\label{fig:phi}
\end{figure}

Figure~\ref{fig:phi} shows an example of application to chromosome 1 based on
data from the case study discussed in Section~\ref{sec:study}. We note
that the mean squared error criterion places more importance on fitting
relatively larger magnitudes close to the diagonal of the image matrix, and so
there is little harm in choosing a moderate value for $\phi$ that best fits the magnitudes of dense groups of correlated SNPs in close proximity.
\end{enumerate}

\subsection{Selecting $\xi_0$ and $\xi_1$}
According to the centroid estimator in~\eqref{eq:centroid}, the $j$-th SNP is
identified as associated if $\pi_j \geq {(1 + \gamma)}^{-1}$.
Following a similar criterion, but with respect to the conditional posteriors,
we have $\Pr(\theta_j = 1 \given \mathbf{y}, \beta, \sigma^2) =
\langle\theta_j\rangle \geq {(1 + \gamma)}^{-1}$, and so,
using~\eqref{eq:etheta},
\[
\logit\langle\theta_j\rangle = -\dfrac{1}{2}\log\kappa + \xi_0 + \xi_1
\mathbf{w}_j^\top \mathbf{r} + \dfrac{\beta_j^2}{2\sigma^2}
\Bigg(1 - \dfrac{1}{\kappa}\Bigg) \geq -\log\gamma.
\]
After some rearrangements, we see that, in terms of $\beta_j$, this criterion
is equivalent to $\beta_j^2 \geq \sigma^2 s_j^2$ with
\begin{equation}
\label{eq:betacentroid}
s_j^2 \doteq \dfrac{2\kappa}{\kappa - 1}
\Bigg(\dfrac{1}{2}\log\kappa - \xi_0 -\xi_1 \mathbf{w}_j^\top\mathbf{r}
-\log\gamma\Bigg),
\end{equation}
that is, we select the $j$-th marker if $\beta_j$ is more than $s_j$ ``spike''
standard deviations $\sigma$ away from zero.

This interpretation based on the EM formulation leads to a meaningful
criterion for defining $\xi_0$ and $\xi_1$: we just require that
$\min_{j=1,\ldots,p} s_j^2 \geq s^2$,
that is, that the smallest number of standard deviations is at least $s > 0$.
Since $\max_{j=1,\ldots,p} \mathbf{w}_j^\top \mathbf{r} = 1$,
\[
\min_{j=1,\ldots,p} s_j^2 = 
\dfrac{2\kappa}{\kappa - 1}
\Bigg(\dfrac{1}{2}\log\kappa - \xi_0 -\xi_1 -\log\gamma\Bigg) \geq s^2,
\]
and so,
\begin{equation}
\label{eq:xi1criterion}
\xi_1 \leq \dfrac{1}{2}\log\kappa - \xi_0 - \log\gamma
-\dfrac{s^2}{2}\Bigg(1 - \dfrac{1}{\kappa}\Bigg).
\end{equation}
For a more stringent criterion, we can take the minimum over $\kappa$ in the
right-hand side of~\eqref{eq:xi1criterion} by setting $\kappa = s^2$.
When setting $\xi_1$ it is also important to keep in mind that $\xi_1$ is the
largest allowable gene boost, or better, increase in the log-odds of a marker
being associated to the trait.

Since $\xi_0$ is related to the prior probability of a SNP being
associated, we can take $\xi_0$ to be simply the logit of the fraction of
markers that we expect to be associated \emph{a priori}. However,
for consistency, since we want $\xi_1 \geq 0$, we also require that the right
hand side of~\eqref{eq:xi1criterion} be non-negative, and so
\begin{equation}
\label{eq:xi0criterion}
\xi_0 + \log\gamma \leq \dfrac{1}{2}\log\kappa
-\dfrac{s^2}{2}\Bigg(1 - \dfrac{1}{\kappa}\Bigg).
\end{equation}
Equation~\eqref{eq:xi0criterion} constraints $\xi_0$ and $\gamma$ jointly, but
we note that the two parameters have different uses: $\xi_0$ captures our
prior belief on the probability of association and is thus part of the model
specification, while $\gamma$ defines the sensitivity-specificity trade-off
that is used to identify associated markers, and is thus related to model
inference.

As an example, if $\gamma = 1$ and we set $s = 4$, then the bound
in~\eqref{eq:xi1criterion} with $\kappa = s^2$ is
$\log(s^2)/2 - s^2(1 - 1/s^2)/2 = -6.11$. If we expect $1$ in $10,\!000$
markers to be associated, we have $\xi_0 = \logit(10^{-4}) = -9.21 <
-6.11$ and the bound~\eqref{eq:xi0criterion} is respected.
The upper bound for $\xi_1$ in~\eqref{eq:xi1criterion} is thus $3.10$.

\subsection{Selecting $\kappa$}
We propose using a metric similar to the Bayesian false discovery rate [BFDR,
\citep{whittemore2007bayesian}] to select $\kappa$.
The BFDR of an estimator is computed by taking the expected value of the
false discovery proportion under the marginal posterior distribution of
$\theta$:
\[
\text{BFDR}(\hat{\theta})
= \Exp_{\theta \given \mathbf{y}} \Bigg[
\frac{\sum_{j=1}^p \hat{\theta}_j (1 - \theta_j)}
{\sum_{j=1}^p \hat{\theta}_j}
\Bigg]
= \frac{\sum_{j=1}^p \hat{\theta}_j (1 - \pi_j)}
{\sum_{j=1}^p \hat{\theta}_j}.
\]

Since, as in the previous section, we cannot obtain estimates of
$\Pr(\theta_j = 1 \given \mathbf{y})$ just by running our EM algorithm, we
consider instead an alternative metric that uses the conditional posterior
probabilities of association given the fitted parameters,
$\langle\theta_j\rangle = \Pr(\theta_j = 1 \given \mathbf{y},
\hat{\beta}_{EM}, \hat{\sigma}^2_{EM})$. We call this new metric EMBFDR:\@
\[
\text{EMBFDR}(\hat{\theta})
= \frac{\sum_{j=1}^p \hat{\theta}_j (1 - \langle\theta_j\rangle)}
{\sum_{j=1}^p \hat{\theta}_j}.
\]
Moreover, by the definition of the centroid estimator in~\eqref{eq:centroid},
we can parameterize the centroid EMBFDR using $\gamma$:
\[
\text{EMBFDR}(\hat{\theta}_C(\gamma)) =
\text{EMBFDR}(\gamma)
= \frac{\sum_{j=1}^p I[\langle\theta_j\rangle \geq {(1+\gamma)}^{-1}]
(1 - \langle\theta_j\rangle)}
{\sum_{j=1}^p I[\langle\theta_j\rangle \geq {(1+\gamma)}^{-1}]}.
\]

\begin{figure}
\begin{center}
\includegraphics[width=.95\textwidth]{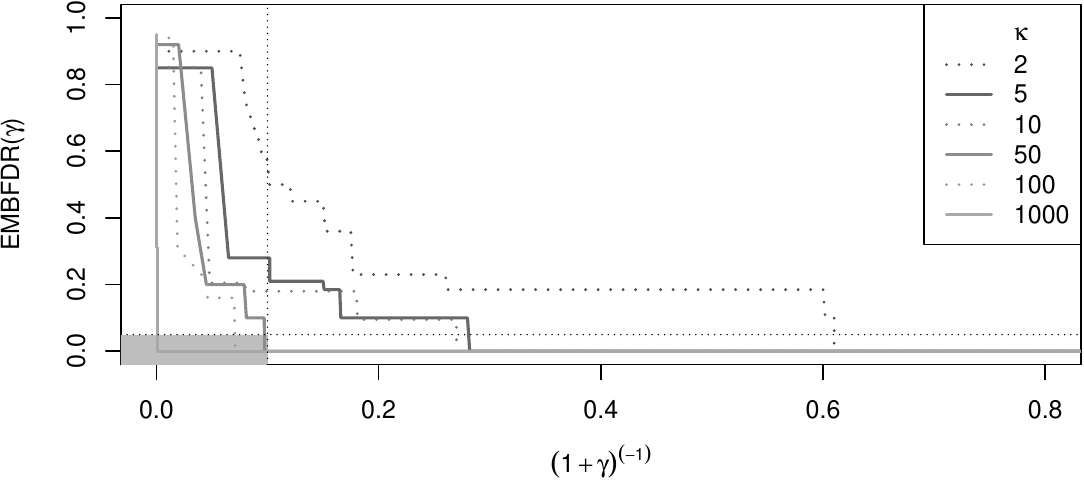}
\caption{When analyzing a data set generated for a simulation study as
described in Section~\ref{sec:study}, we inspect the behavior of the BFDR as a
function of $\gamma$ for various values of $\kappa$ and see that a choice of
$\kappa = 1,000$ would be appropriate to achieve a BFDR no greater than 0.05
when using a threshold of ${(1 + \gamma)}^{-1} = 0.1$.}
\label{fig:bfdrkappa}
\end{center}
\end{figure}

We can now analyze a particular data set using a range of values for $\kappa$
and subsequently make plots of the EMBFDR metric as a function of the
threshold ${(1 + \gamma)}^{-1}$ or as a function of the proportion of SNPs
retained after the EM filter step. Thus, by setting an upper bound for a desired
value of the EMBFDR we can investigate these plots and determine an
appropriate choice of $\kappa$ and an appropriate range of values of $\gamma$.
In Figure~\ref{fig:bfdrkappa} we illustrate an application of this criterion.
We note that the EMBFDR has broader application to Bayesian variable selection
models and can be a useful metric to guide the selection of tuning parameters,
in particular the spike-and-slab variance separation parameter $\kappa$.

\subsection{Visualizing the relationship between SNPs and genes}
\begin{figure}[htbp]
\begin{center}
\includegraphics[width=.95\textwidth]{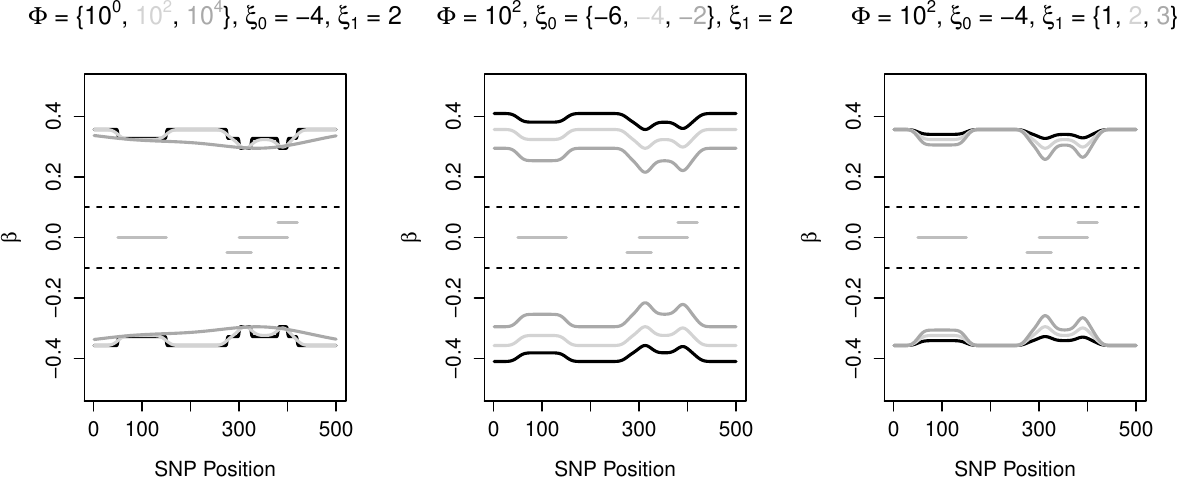}
\caption{We illustrate the effect of varying $\phi$, $\xi_0$ and
$\xi_1$ on the thresholds on the posterior effect sizes, $\beta_j$, in a
simple window containing a single gene in isolation, and a group of three
overlapping genes. On the left, we vary $\phi$ and control the smoothness of
the thresholds. In the middle, we vary $\xi_0$ and control the magnitude of
the thresholds, or in other words the number of standard deviations ($\sigma$)
away from zero at which they are placed. On the right, we vary $\xi_1$ and
control the sharpness of the difference in the thresholds between differently
weighted regions of the window. For this illustration, we set $\sigma^2 =
0.01$, $\kappa = 100$, and $\gamma = 1$. We mark the distance $\sigma$ away
from the origin with black dashed lines.}
\label{fig:paramvis}
\end{center}
\end{figure}

For a given configuration of $\kappa$, $\gamma$, and $\sigma^2$, we can plot
the bounds $\pm\sigma s_j$ on $\beta_j$ and inspect the effect of parameters
$\phi$, $\xi_0$, and $\xi_1$. SNPs that are close to relevant genes have
thresholds that are relatively lower in magnitude; they need a relatively
smaller (in magnitude) coefficient to be selected for the final model. With
everything else held fixed, as $\phi \to \infty$ the boost received from the
relevant genes will decrease to zero and our model will coincide with a basic
version of Bayesian variable selection where
$\theta_j \iid \text{\sf Bern}(\logit^{-1}(\xi_0))$. We demonstrate this
visualization on a mock chromosome in Figure~\ref{fig:paramvis}.

\section{Empirical Studies}
\label{sec:study}
We conduct two simulation studies. First, we compare the performance of our
method to other well-known methods including single SNP tests, LASSO, fused
LASSO, group LASSO, the penalized unified multiple-locus association (PUMA)
suite of~\cite{hoffman2013puma}, and the Bayesian sparse linear mixed
model (BSLMM) of~\cite{zhou2013polygenic}. Then we assess the robustness of our
method to misspecifications of the range parameter, $\phi$, and gene
relevances. We describe each study in detail below, but we first explain how
the data is simulated in each scenario.

\subsection{Simulation Study Details}
\label{sec:ssdetails}
To provide a fair comparison across methods and to realistically assess
the robustness of our method to misspecifications, we designed our simulation
study based on real-life genotypical data and current gene and marker
annotations. Specifically, to keep a representative LD structure we sample
whole-chromosome individual genotypes by subsampling individual data provided by
the~\citet{10002012integrated}; gene weights are computed based on gene
lengths and positions in the hg19 reference, while marker positions are taken
from actual SNP array designs in the WTCCC studies. We consider two studies: a
``non-informative'' setup where the gene relevances are uniformly set to one
and $\phi = 10^8$, so that marker relevance scores ${\mathbf{w}_j(\phi)}^\top
\mathbf{r}$ are small and close to uniform, and a mild boost effect of $\xi_1
= 1$; and an ``informative'' study with gene relevances taken as search hit
scores from~\citet{MalaCards} and $\phi = 10^4$, a frequent value when
adopting the procedure in Section~\ref{sec:phi}, and a stronger boost effect,
$\xi_1 = 5$. These two studies are extreme with respect to marker relevance
scores---a function of gene relevances and genomic range---and spatial boost
effects and aim at assessing how robust are our model and recommended
guidelines. For instance, when fitting all scenarios we take an informative
approach by considering the same gene relevances and genomic range from the
``informative'' scenario, but adopt a conservative approach by setting
$\xi_1 = 1$ as in the ``non-informative'' scenario.

For both studies we simulated two scenarios for the number of markers $p$: the
``small'' scenario comprised chromosome~19 with $p = 4,\!199$ markers, and the
``large'' scenario containing all $p = 28,\!932$ markers in chromosome~2.
Chromosomes~19 and~2 are the smallest and largest in terms of number of
markers, respectively. We kept the ratio of $p$ to the number of individuals
$n$ at $50$, representative of real-life studies, so $n = 85$ in the small
scenario and $n = 580$ in the large scenario. In all simulations we fix the
number of causal markers $m = 10$ and set the baseline log-odds
$\xi_0 = \lfloor \logit(m / p) \rfloor$. For each simulation batch, we first
sample uniformly at random $n$ individuals from the 1000 Genomes dataset,
taking their whole chromosome genotypes according to the small or large
scenario, and filter out markers with sampled MAF $<0.05$, deemed as rare
variants, or $>0.50$. Next, we sample $m$ causal markers
following~\eqref{eq:thetaprior}, with marker relevance scores and $\xi_1$
taken according to a non-informative or informative scenario. Effect sizes
$\beta_j$ are then sampled to reflect the challenging nature of GWAS:\@
$\beta_j \given \theta_j \sim \theta_j N(0, 0.25) + (1 - \theta_j) N(0,
0.01)$, that is, small effect sizes for causal markers and relatively large
coefficients for noisier non-causal effects. Finally, for each replicate
within a batch we sample phenotypes according to~\eqref{eq:lhood}.

In each simulation scenario and dataset below we fit the model as follows: we
adopt informative gene relevances from MalaCards and $\phi = 10,\!000$, start
with conservative values for the baseline log-odds $\xi_0 = \logit(100 / p)$
and the gene boost effect $\xi_1 = 1$, and run the EM filtering process until
either the predictive performance starts degrading or at most $10$ markers
remain. We measure predictive performance using a metric similar to posterior
predictive loss~\citep[PPL;][]{gelfand1998model}: if, at the $t$-th EM
iteration,
$\hat{y}^{(t)}_i = \Exp[y_{i,\text{rep}} \given \hat{\beta}^{(t)}_{\text{EM}},
\mathbf{y}]$ is the $i$-th predicted response, the PPL measure under squared
error loss is approximated by
\[
\text{PPL}(t) = \sum_{i=1}^n {(y_i - \hat{y}_i)}^2 + \sum_{i=1}^n
\text{Var}[y_{i,\mbox{\tiny rep}} \given \hat{\beta}^{(t)}_{EM}, \mathbf{y}]
= \sum_{i=1}^n {(y_i - \hat{y}_i)}^2 + \hat{y}_i(1 - \hat{y}_i).
\]
The right panel in Figure~\ref{fig:rocppl} shows an example of how the
relative PPL (rPPL) typically varies as the EM filter advances. We define the
rPPL at the $t$-th EM iteration as the ratio between PPL$(t)$ and the
PPL of the null model, that is, the model with only the intercept and no
marker genotypes as predictors. At the end of the filtering stage we run the
Gibbs sampler with $\xi_0 = \logit(m / \tilde{p})$, where $\tilde{p}$ is the
number of markers retained at the end of the EM filter. Parameter $\kappa$ is
actually elicited at each EM filtering iteration using EMBFDR.\@

\subsection{Comparison Simulation Study}
In this study, we generated 10 batches of simulated data, each containing 5
replicates, for a total of 50 simulated data sets for each cross configuration
of small and large scenarios by non-informative and informative studies.
After simulating the data, we fit our model and compared its performance in
terms of area under the receiver operating characteristic (ROC) curve (AUC) to
the usual single SNP tests, LASSO, fused LASSO, group LASSO, PUMA, and BSLMM
methods. We used the \textbf{penalized} package in \textsf{R} to fit the LASSO
and fused LASSO models; we used ten-fold cross-validation to determine the
optimal values for the penalty terms. Similarly, we used the \textbf{gglasso}
package in \textsf{R} to fit the group LASSO model where we defined the groups
such that any two adjacent SNPs belonged to the same group if they were within
$10,\!000$ base pairs of each other; we used ten-fold cross validation to
determine the optimal value for the penalty term. Finally, we used the
authors' respective software packages to fit the PUMA and BSLMM models.

To calculate the AUC for any one of these methods, we took a final ranking of
SNPs based on an appropriate criterion (see more about this below), determined
the points on the ROC curve using our knowledge of the true positives and the
false positives from the simulated data, and then calculated the area under
this curve. For our model, we used either the ranking (in descending order) of
$\mathbb{E}[\theta_j \given \hat{\beta}_{\text{EM}},
\hat{\sigma}^2_{\text{EM}}, y]$
for a particular EM filtering step or
$\hat{\mathbb{P}}(\theta_j = 1 \given y)$ using the samples obtained by the
Gibbs sampler; for the single SNP tests we used the ranking (in ascending
order) of the p-values for each marker's test; for LASSO, fused LASSO and
group LASSO we used the ranking (in descending order) of the magnitude of the
effect sizes of the SNPs in the final model; for the other penalized
regression models given by the PUMA program, we used the provided software to
compute p-values for each SNP's significance in the final model and used the
ranking (in ascending order) of these p-values; for BSLMM we used the ranking
(in descending order) of the final estimated posterior probabilities of
inclusion for each SNP in the final model.

\begin{figure}
\begin{center}
\includegraphics[width=0.48\textwidth]{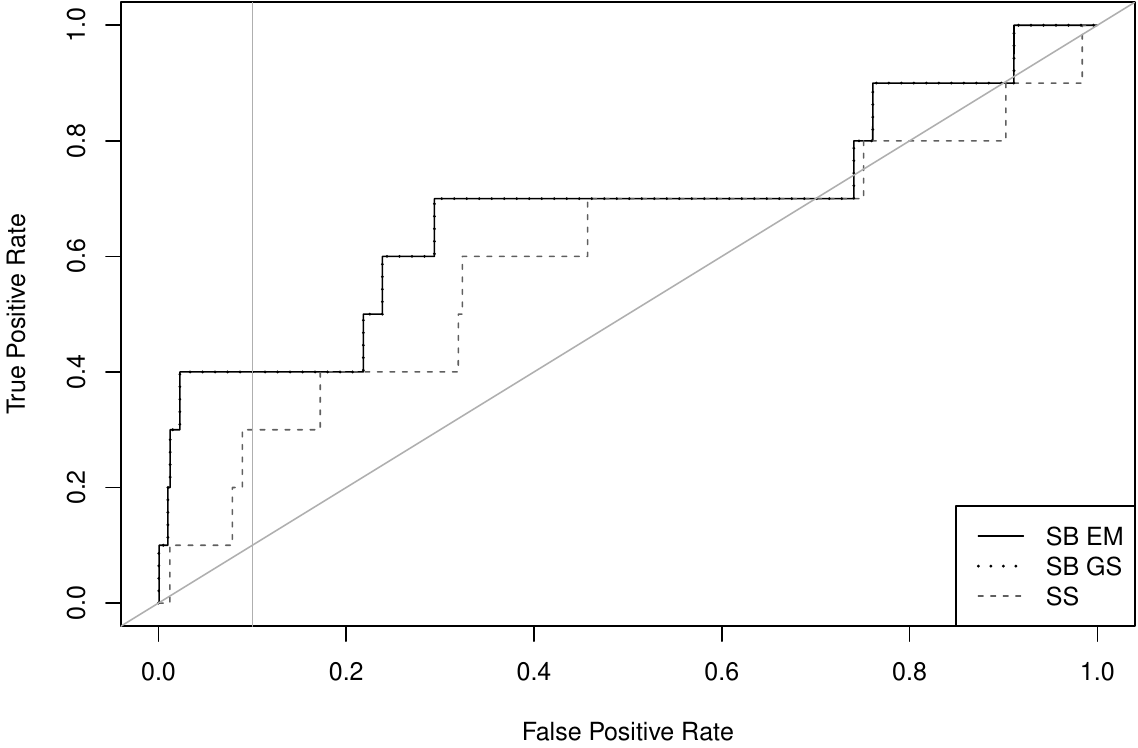}
\includegraphics[width=0.48\textwidth]{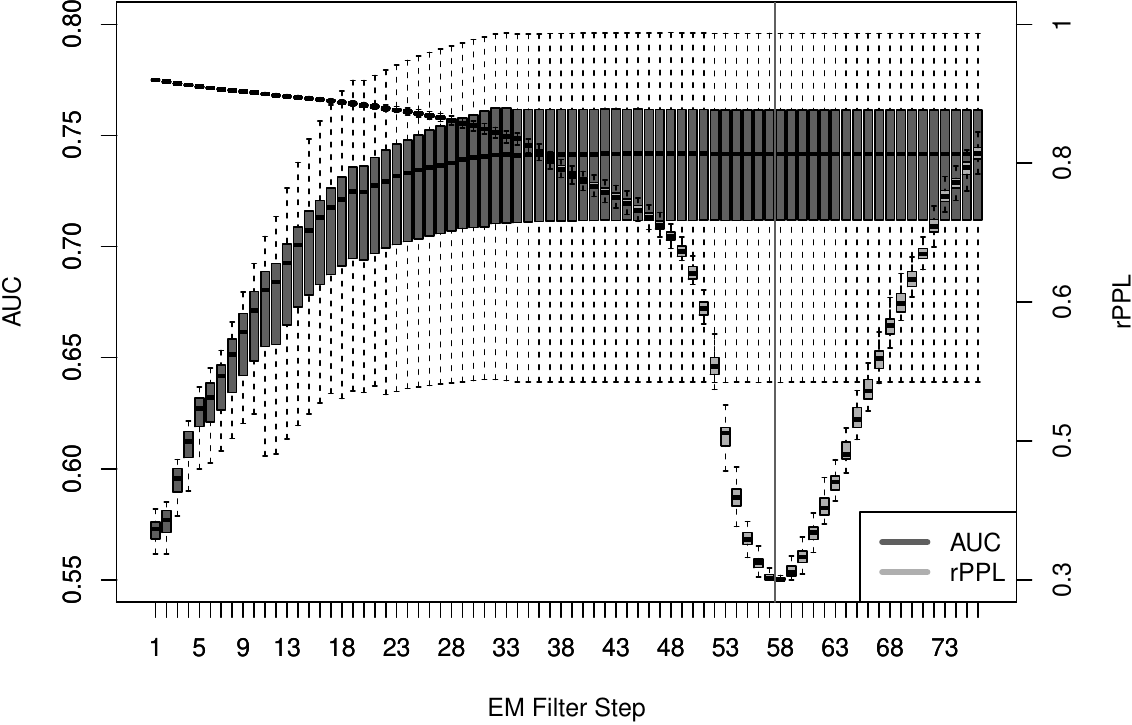}
\end{center}
\caption{Left: examples of ROC curves for our method (solid line for EM
results, crosses for Gibbs sampling results) and single SNP tests (dashed
line); vertical gray line marks a false positive rate of 10\%. Right:
relation between AUC and relative PPL (rPPL) as the EM filter progresses,
boxplots over replicates within a simulation batch; in this case the EM filter
would stop at iteration 58, right before rPPL starts increasing.}
\label{fig:rocppl}
\end{figure}

The left panel in Figure~\ref{fig:rocppl} shows examples of ROC curves from a
random simulation replicate. Since the interest in GWA studies is focused on
low false positive rates, we also evaluate the performance of the methods with
respect to \emph{relative} AUC (rAUC) at some false positive rate $f$, defined
simply as the normalized AUC up to FPR $f$, that is, rAUC$(f)$ = AUC$(f) / f$.
According to this criterion, the solid ROC curve in Figure~\ref{fig:rocppl}
has a clearer advantage over the dashed curve at a 10\% FPR.\@ The right panel
illustrates how the AUC changes with rPPL and as the EM filtering iterations
increase.

\begin{figure}
\begin{center}
\includegraphics[width=0.47\textwidth]{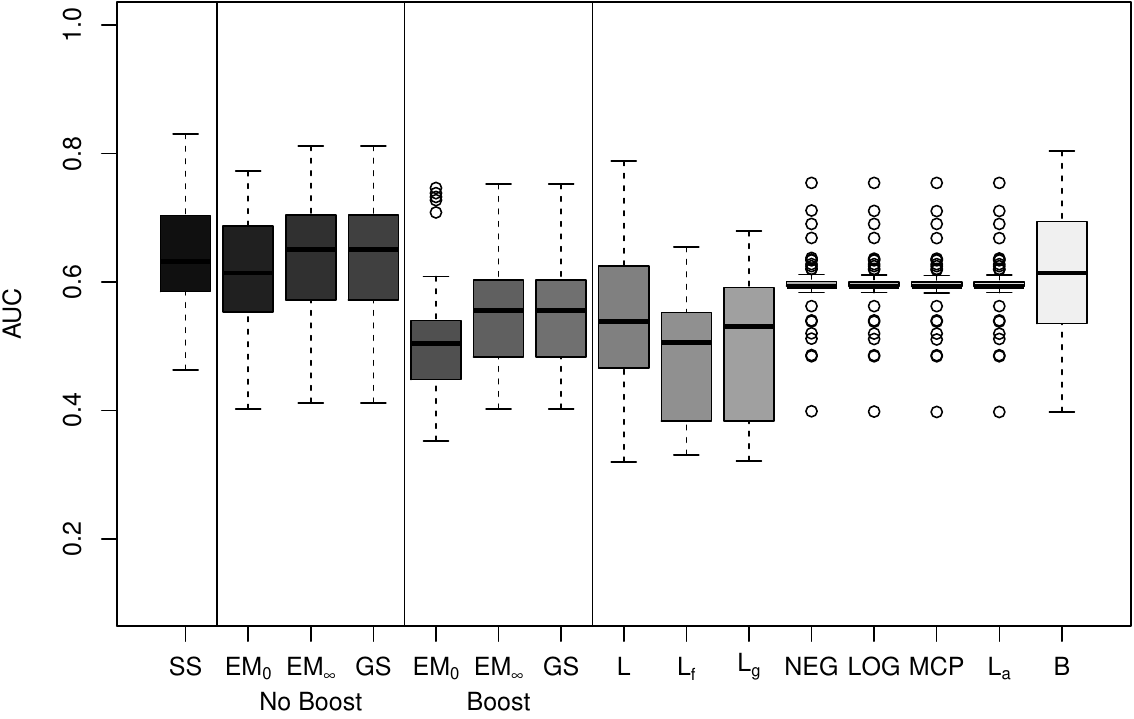}
\includegraphics[width=0.47\textwidth]{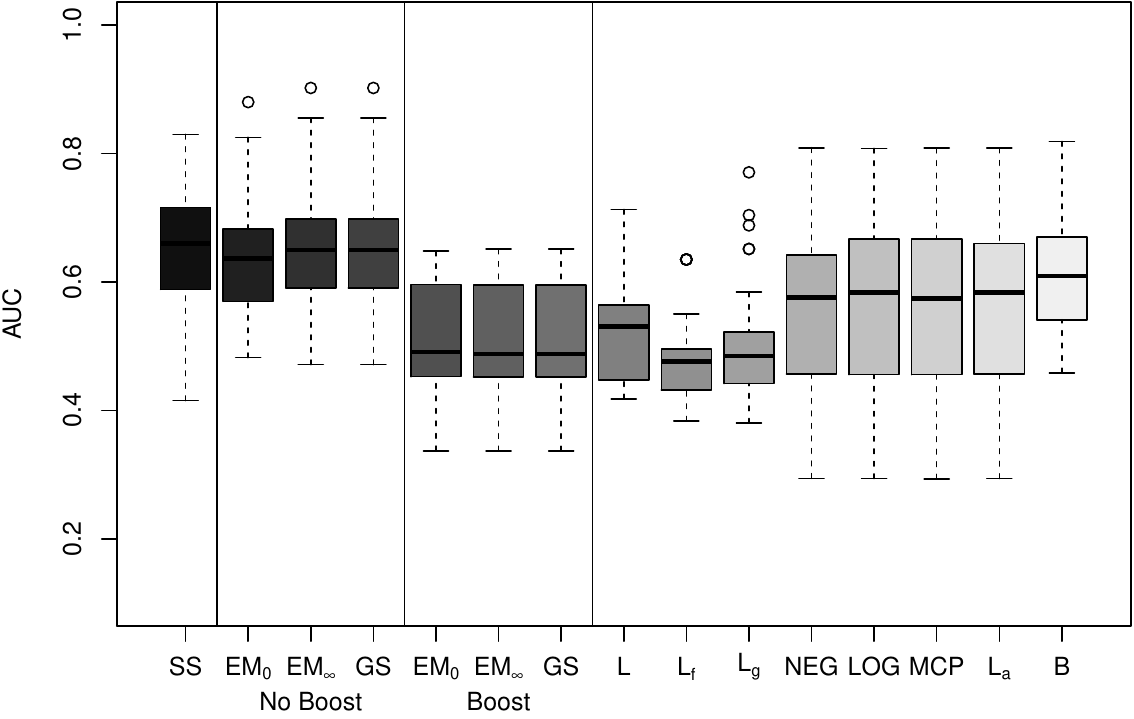}

%\vspace{1em}

%\includegraphics[width=0.48\textwidth]{simstudy2-chr19-inf-crop.pdf}
%\includegraphics[width=0.48\textwidth]{simstudy2-chr2-inf-crop.pdf}
\includegraphics[width=0.47\textwidth]{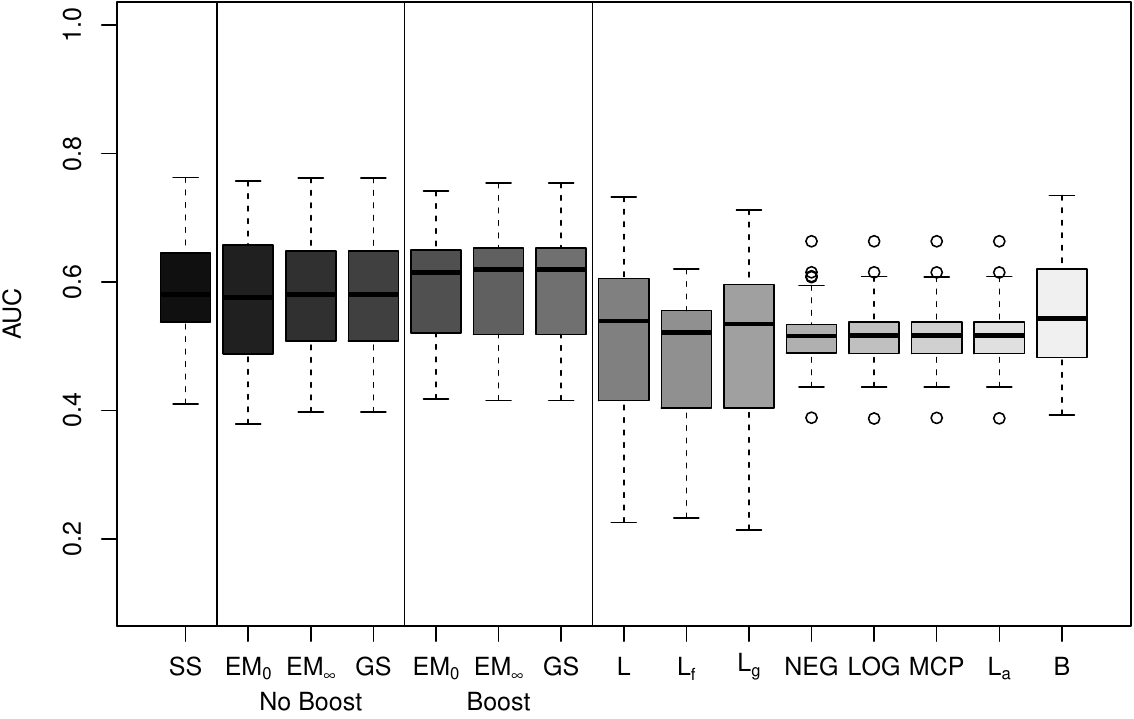}
\includegraphics[width=0.47\textwidth]{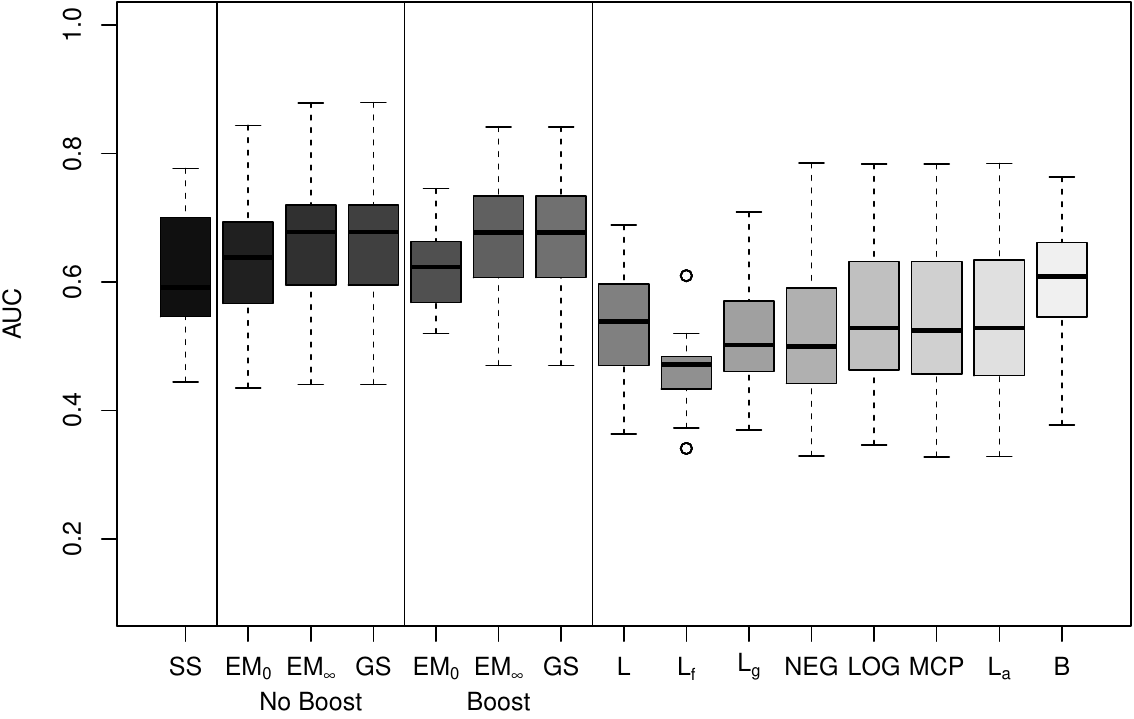}

\vspace{1em}

\includegraphics[width=0.47\textwidth]{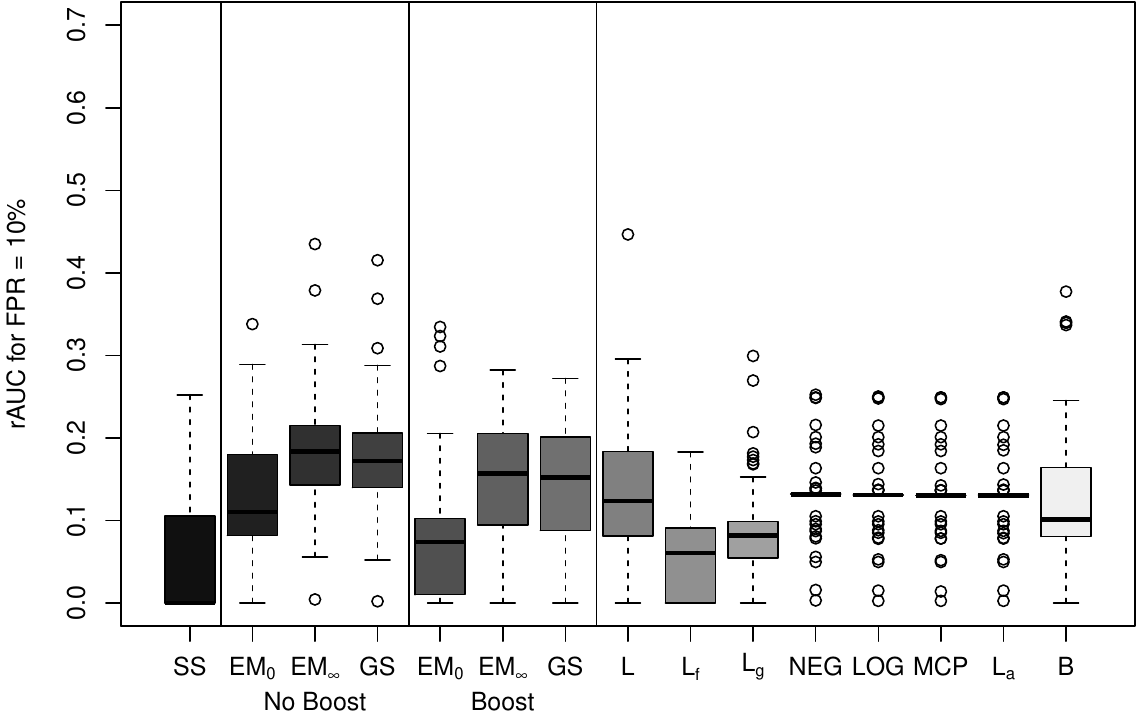}
\includegraphics[width=0.47\textwidth]{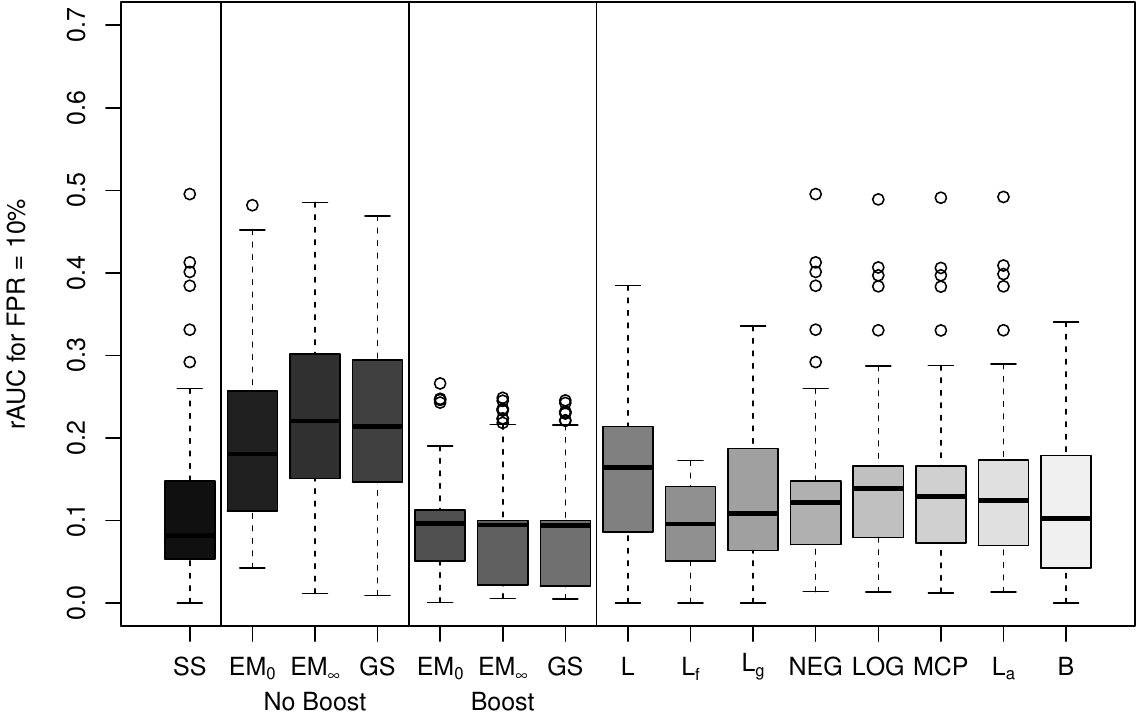}

%\vspace{1em}

\includegraphics[width=0.47\textwidth]{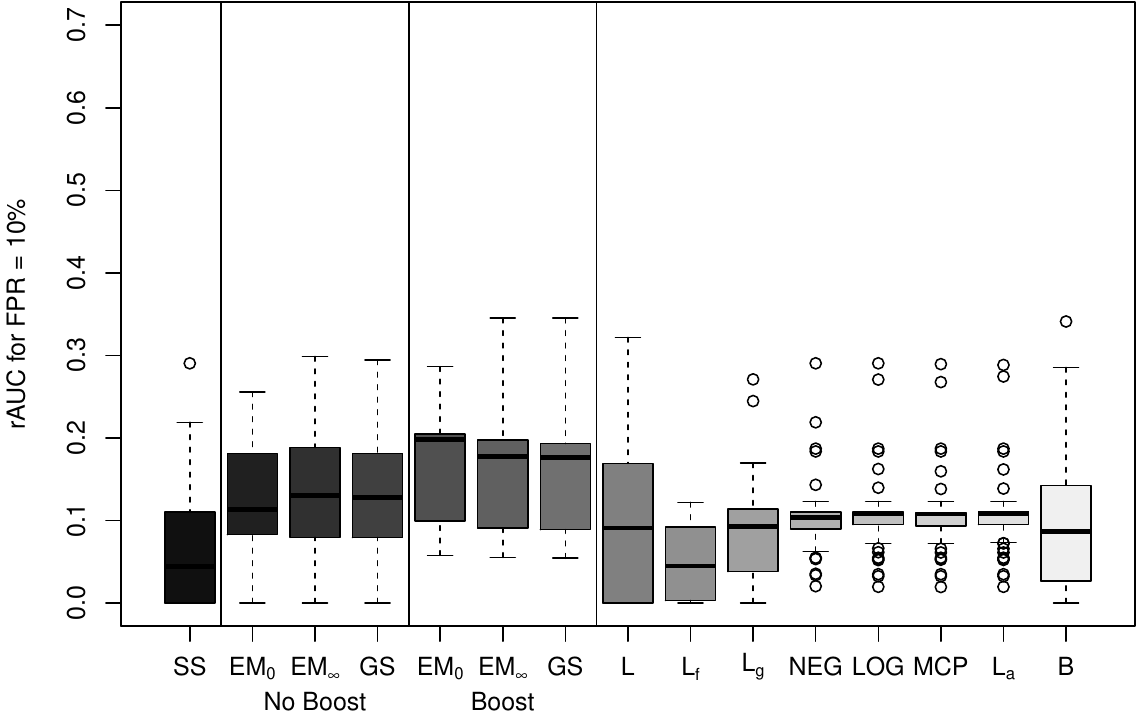}
\includegraphics[width=0.47\textwidth]{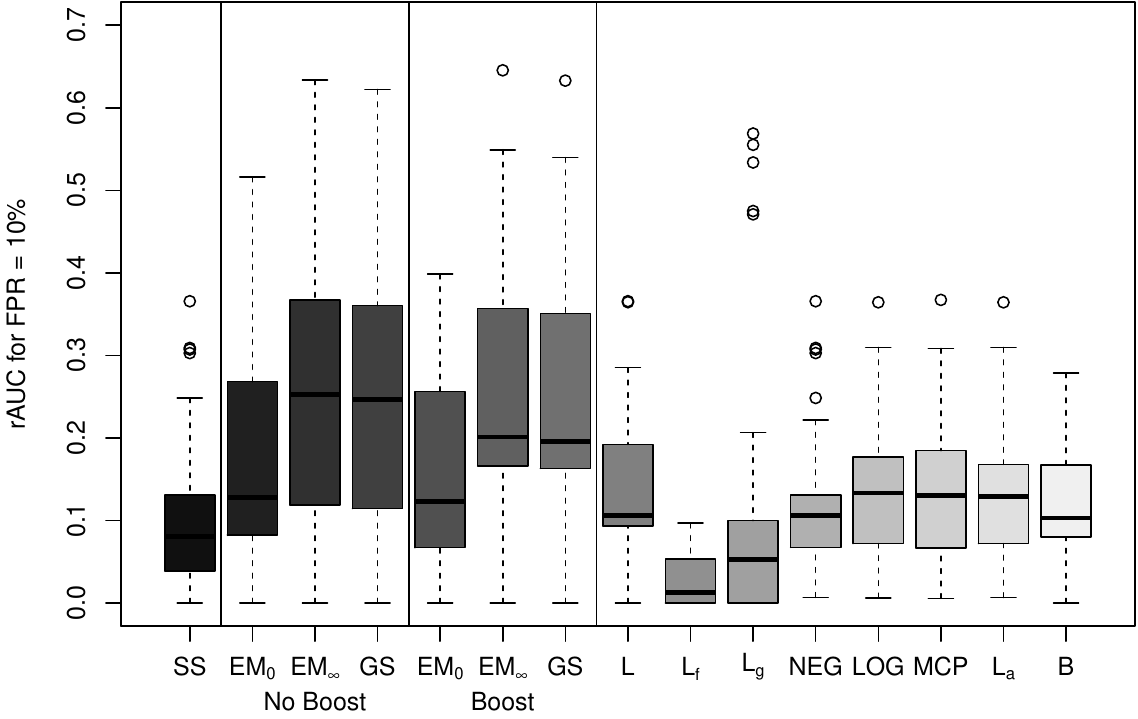}
\end{center}
\caption{Results from the comparison simulation study. Top four panels show
AUC, while bottom four panels show relative AUC.\@ Within each block of four
panels: left panels show results under non-informative scenario (top) and
informative scenario (bottom) for small study, while right panels show
respective AUC or rAUC results for the large study. The boxplots in each panel
are, left to right: single SNP (SS) tests; spatial boost ``null'' (no boost)
model at the first (EM$_0$) and last (EM$_\infty$) iteration of the EM filter
and after Gibbs sampling phase; spatial boost ``informative'' model at first
and last EM filtering iterations and after Gibbs sampling phase; LASSO (L);
fused LASSO (L$_f$); grouped LASSO (L$_g$); PUMA with penalties NEG, LOG, MCP,
and adaptive LASSO (L$_a$); and BSLMM (B).}
\label{fig:comparison}
\end{figure}

We summarize the results in Figure~\ref{fig:comparison}. With respect to AUC
(top four panels), our methods perform better on average than the other
methods in the informative scenario, but only comparably in the
non-informative scenario. Note that all models have a modest performance due
to the challenging nature of the simulation (but our model has improved
performance under less stringent scenarios; see~\ref{suppA} for more
details). Most of the gain in our methods comes at the beginning of the ROC
curves, i.e.\ at low false positive rates, as exemplified in
Figure~\ref{fig:rocppl}. This becomes more evident if we compare relative AUC
in the bottom four panels. We note that the gains are
present even in the ``null'' models with $\xi_1 = 0$, so they stem from the
joint modeling of markers. Additional gains in rAUC are more apparent in the
informative scenario (bottom row).

The seemingly bad performance of our model in the non-informative scenario
indicates that the model can be sensitive to poor marker relevance scores,
arising either from meaningless gene relevances or a non-representative
genomic range $\phi$. This is not surprising given that the inference relies
on good prior information in the large-$p$-small-$n$ regimen; however, as the
informative scenario suggests, the model is more robust to misspecification of
$\xi_1$, which can be seen as the overall prior strength of relevance scores.
The null model, for instance, often offers the best performances in both AUC
and relative AUC, which recommends conservative, low values for $\xi_1$ in
initial analyses. Moreover, the EM filtering procedure contributes to further
gains in performance even in non-informative scenarios, that is, with
misspecified relevance scores. These gains are even more pronounced with
larger causal effect sizes relative to non-causal effects, as we show in~\ref{suppA}; that is, we observe that the EM filter becomes more robust to
these misspecifications, especially on large scenarios and under low false
positive rates, as shown here.

\subsection{Relevance Robustness Simulation Study}
To investigate how robust our model is to misspecifications of gene relevances
and genomic range, we focus on their joint effect in defining marker relevance
scores and again consider the small (chromosome~19) and large (chromosome~2)
scenarios under the informative study, where the gene relevances are more
varied. We randomly select one batch from the comparison study to be the
reference in each scenario, with its replicates serving as ground truth. We
then vary three sampling percentages $\pi \in \{25\%, 50\%, 75\%\}$, and, for
each $\pi$, we randomly select $\lfloor \pi p \rfloor$ markers uniformly from
each reference replicate and sample their relevance scores from the empirical
distribution of relevance scores in each scenario. This simulation is
replicated 50 times.

\begin{figure}
\centering
\begin{minipage}{.2\textwidth}
  \centering
  \includegraphics[width=\textwidth]{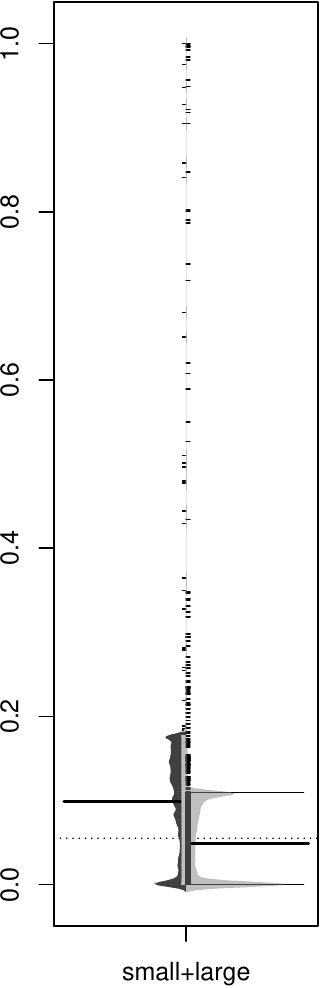}
\end{minipage}\hspace{.04\textwidth}%
\begin{minipage}{.76\textwidth}
  \centering
  \includegraphics[width=\textwidth]{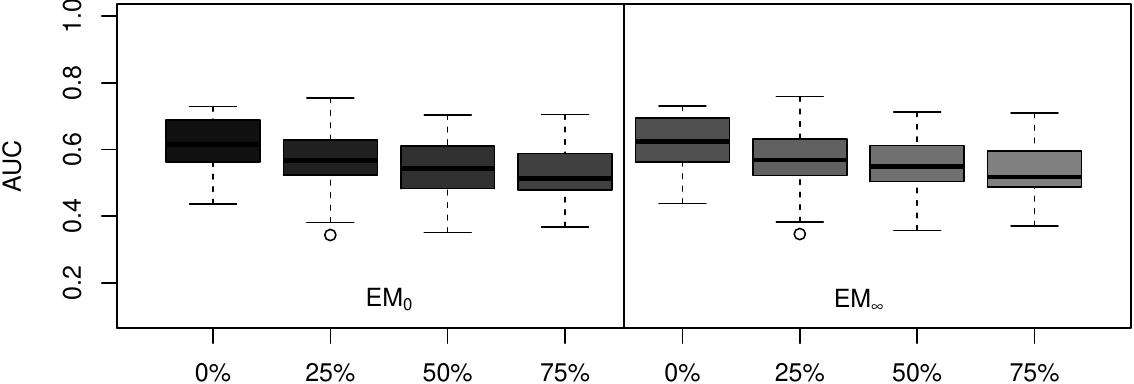}

  \vspace{2em}

  \includegraphics[width=\textwidth]{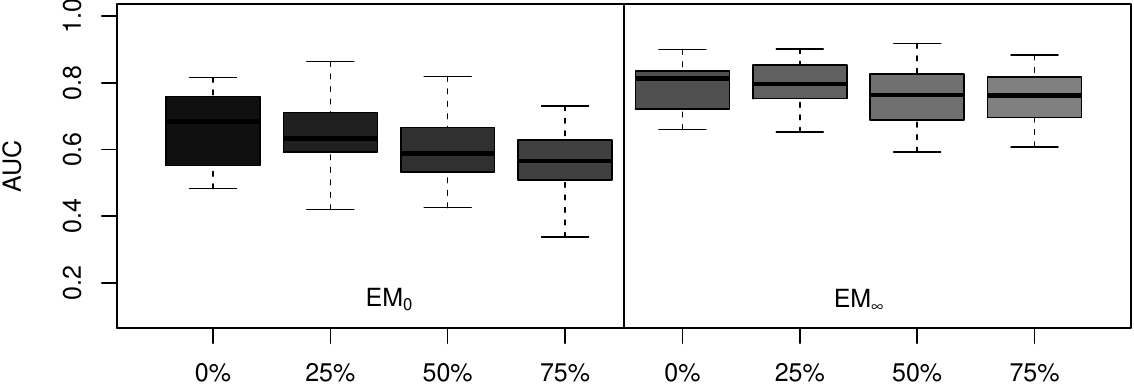}
\end{minipage}
\caption{Results from the simulation study to assess robustness to marker
relevance scores. Left panel shows the distribution of marker relevance score
from small and large scenarios, chromosomes~19 and~2, respectively.
Right panels show AUC results for first (EM$_0$) and last (EM$_\infty$)
filtering iterations across simulation percentages for small (top) and large
(bottom) scenarios. Null percentages correspond to reference replicates.}
\label{fig:robustness}
\end{figure}

Hyper-prior parameters $\xi_0$, $\xi_1$, and $\kappa$ were elicited as in
Section~\ref{sec:ssdetails}. For each simulated replication we then fit our
model and assess performance using the AUC, as in the previous study.
Figure~\ref{fig:robustness} illustrates the distribution of relevance scores
for the markers in chromosomes~19 and~2 and how the performance of the model
varies at each sample percentage. The AUC at initial EM iterations degrades
with higher percentages for both scenarios, as expected; however, small
scenario replicates continue to show a similar pattern at their last EM
iterations, as opposed to large scenario replicates that show better and
stable results across all percentages. We attribute this discrepancy in
robustness to the distribution of relevance scores. As the left panel in
Figure~\ref{fig:robustness} shows, the large scenario has a bimodal
distribution and low mean score and so a few markers are relevant; in
contrast, the small scenario score distribution has more spread and higher
mean and so many markers can be relevant and influence negatively the fit by
calling more false positives. We thus recommend to analyze the resulting
distribution of marker relevance scores when eliciting gene relevances and
$\phi$.

% pdf(file='rel-scores.pdf', width=3, height=8, onefile=T)
% beanplot(w19, w2, border=NA, log="", side="both", ll=.03, col=list(c("gray25", "gray75"), c("gray75", "gray25")), names=c('small', 'large'))

\section{Case Study}
\label{sec:case}
Using data provided by the WTCCC, we analyzed the entire genome (342,502
SNPs total) from a case group of~1,999 individuals with rheumatoid arthritis
(RA) and a control group of~1,504 individuals from the~1958 National Blood
Bank dataset. For now we addressed the issues of LD, rare variants, and
population stratification by analyzing only the SNPs in Hardy-Weinberg
Equilibrium~\citep{wigginton2005note} with minor allele frequency greater than
5\%. There are 15 SNPs that achieve genome-wide significance when using a
Bonferroni multiple testing procedure on the results from a single SNP
analysis. Table 1 in~\ref{suppA} provides a summary of these results for
comparison to those obtained when using the spatial boost model. 

When fitting the spatial boost model, we broke each chromosome into blocks and
selected an optimal value of $\phi$ for each block using our proposed method
metric, $|\rho_{i,j}|(\phi)$. We used the EMBFDR to select a choice for
$\kappa$ from the set $\{10^2, 10^3, 10^4, 10^5, 10^6\}$ at each
step of our model fitting pipeline so that the BFDR was no greater than 0.05
while retaining no larger than 5\% of the total number of SNPs. With a
generous minimum standard deviation $s = 1$ we have that trivially $\xi_0 < 0$
from~\eqref{eq:xi0criterion}, but we set $\xi_0 = -8$ to encode a prior belief
that around 100 markers would be associated to the trait on average \emph{a
priori}. The bound on $\xi_1$ is then $\xi_1 \le 8$, but we consider log
odds-ratio boost effects of $\xi_1 \in \{1, 4, 8\}$. A value of $\xi_1 = 1$ is
more representative of low power GWA studies; however, the larger boost
effects offer more weight to our prior information.
For comparison, we also fit a model without any gene boost by setting $\xi_1 =
0$ (the ``null'' model), and also fit two models for each possible value of
$\xi_1$ trying both a non-informative gene relevance vector and a vector based
on text-mining scores obtained from \citet{MalaCards}.

\begin{figure}%[!h]
\begin{center}
\includegraphics[width=.95\textwidth]{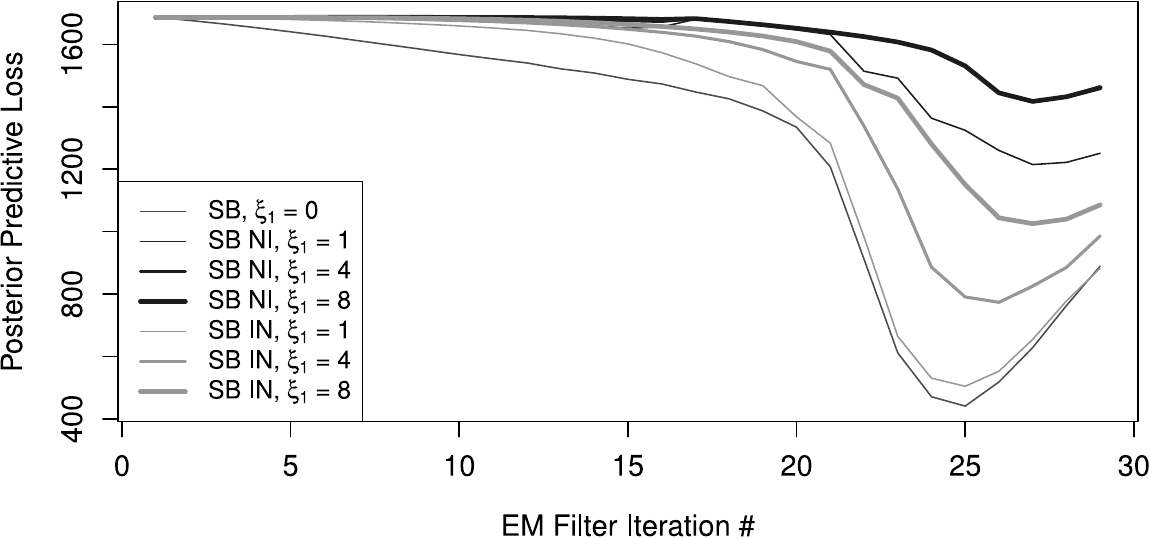}
\caption{Although we run the EM filter until the number of retained markers
$<$ 100 (iteration \#29), the PPL metric often tells us to keep between 200 to
250 markers (iterations \#25--26).}
\label{fig:PPL}
\end{center}
\end{figure}

To accelerate the EM algorithm, we rank-truncate $X$ using $l = 3,\!259$
singular vectors; the mean squared error between $X$ and this approximation is
less than 1\%. We apply the EM filtering 29 times and use PPL to decide when
to start the Gibbs sampler. As Figure~\ref{fig:PPL} shows, in all of our
fitted models, the PPL decreases slowly and uniformly for the first twenty or
so iterations, and then suddenly decreases more sharply for the next five
iterations until it reaches a minimum and then begins increasing uniformly
until the final iteration, similarly to the behavior depicted in
Figure~\ref{fig:rocppl}. For comparison to the 15~SNPs that achieve
genome-wide significance in the single marker tests, Tables 2-15 in~\ref{suppA} list, for
each spatial boost model, the top 15~SNPs at the optimal EM filtering step,
i.e.\ the step with the smallest PPL, and the top 15~SNPs based on the
posterior samples from our Gibbs sampler when using only the corresponding set
of retained markers.

We observe the most overlap with the results of the single SNP tests in our
null model where $\xi_1 = 0$ and in our models that use informative priors
based on relevance scores from MalaCards. Although there is concordance
between these models in terms of the top 15 SNPs, it is noteworthy that we
select only a fraction of these markers after running either the EM algorithm
or the Gibbs sampler. Based on the results from our simulation study where we
observe superior performances for the spatial boost model at low false
positive rates, we believe that an advantage of our method is this ability to
highlight a smaller set of candidate markers for future investigation. 

Indeed, after running our complete analysis, we observe that the usual
threshold of $0.5$ on $\hat{\mathbb{P}}(\theta_j = 1 | y)$ would result in
only the null spatial boost model ($\xi_1 = 0$), the low gene boost
non-informative model ($\xi_1 = 1$), and the informative models selecting SNPs
for inclusion in their respective final models. The SNPs that occur the most
frequently in these final models are the first four top hits from the single
SNP tests: rs4718582, rs10262109, rs6679677, and rs664893. The SNP with the
highest minor allele frequency in this set is rs6679677; this marker has
appeared in several top rankings in the GWAS literature (e.g.
\citep{burton2007genome}) and is in high LD with another SNP in gene PTPN22
which has been linked to RA \citep{michou2007linkage}.

\begin{figure}%[!h]
\begin{center}
\includegraphics[width=0.7\textwidth]{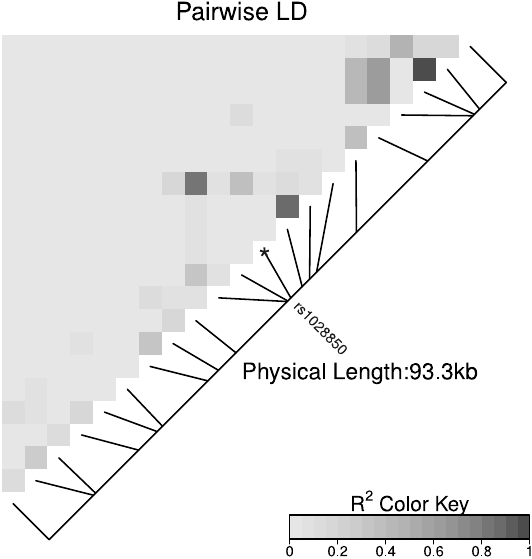}

\vspace{1em}

\includegraphics[width=0.5\textwidth]{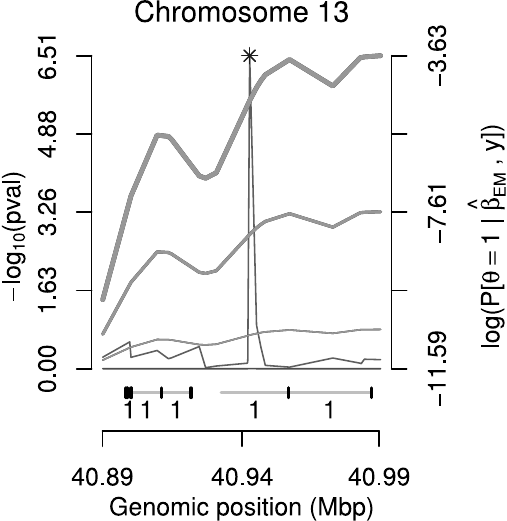}
\hspace{2em}
\includegraphics[width=0.25\textwidth]{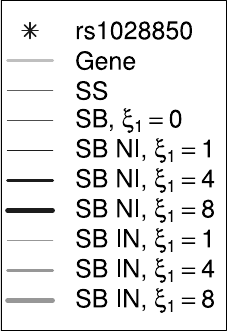}
\caption{Although rs1028850 has a relative peak in the Manhattan plot (SS), it
does not achieve genome-wide significance. The spatial boost (SB) model
initially prioritizes markers that are closer to the center of regions rich in
genes, but selects rs1028850 for inclusion in the final model by the end of
the EM filter (not shown) under several configurations.}
\label{fig:rs1028850}
\end{center}
\end{figure}

If we consider only the final models obtained after running the EM filter, we
see another interesting SNP picked up across the null and informative models:
rs1028850. In Figure~\ref{fig:rs1028850}, we show a closer look at the region
around this marker and compare the trace of the Manhattan plot with the traces
of each spatial boost model's $\mathbb{E}[\theta_j | \hat{\beta}_{EM},
\hat{\sigma}^2_{\text{EM}}, y]$ values at the first iteration of the EM
filter. To the best of our knowledge this marker has not yet been identified
as being associated to RA;\@ moreover, it is located inside a non-protein coding
RNA gene, LINC00598, and is close to another gene that has been linked to RA,
FOXO1 \citep{grabiec2014jnk}.

As we increase the strength of the gene boost term with a non-informative
relevance vector, the relatively strong prior likely leads to a
mis-prioritization of all SNPs that happen to be located in regions rich in
genes. In the supplementary tables (2-15) of~\ref{suppA} we list the lengths of the genes that
contain each SNP and we see that indeed the non-informative gene boost models
tend to retain SNPs that are near large genes that can offer a generous boost.
Perhaps due to prioritizing the SNPs incorrectly in these models, we do not
actually select any markers at either the optimal EM filtering step or after
running the Gibbs sampler. However, some of the highest ranking SNPs for these
models, rs1982126 and rs6969220, are located in gene PTPRN2 which is
interestingly a paralog of PTPN22. 

\section{Conclusions}
\label{sec:conclusion}
We have presented a novel hierarchical Bayesian model for GWAS that exploits
the structure of the genome to define SNP-specific prior distributions for the
model parameters based on proximities to relevant genes.
While it is possible that other ``functional'' regions are also very
relevant---e.g.\ regulatory and highly conserved regions---and that mutations in
SNPs influence regions of the genome much farther away---either upstream,
downstream, or, through a complex interaction of molecular pathways, even 
on different chromosomes entirely---we believe that incorporating information
about the genes in the immediate surroundings of a SNP is a reasonable place
to start.

By incorporating prior information on relevant genomic regions, we focus on
well annotated parts of the genome and were able to identify, in real data,
markers that were previously identified in large studies and highlight at
least one novel SNP that has not been found by other models.
Clearly, validation via other studies of the importance of this
marker for RA should be investigated.
In addition, as shown in a simulation study, while logistic regression under
large-$p$-small-$n$ regimen is challenging, the spatial boost model often
outperforms simpler models that either analyze SNPs independently or employ a
uniform penalty term on the $L_1$ norm of their coefficients.
%Importantly, this better performance is often concentrated at lower false
%positive rates, and thus our findings are of practical relevance.

Our main point is that we regard a fully joint analysis of all markers as
essential to overcome genotype correlations and rare variants. This approach,
however, entails many difficulties. From a statistical point of view, the
problem is severely ill-posed so we rely on informative, meaningful priors to
guide the inference. From a computational perspective, we also have the
daunting task of fitting a large scale logistic regression, but we make it
feasible by reducing the dimension of both data---intrinsically through rank
truncation---and parameters---through EM filtering. Moreover, from a practical
point of view, we provide guidelines for selecting hyper-priors, reducing
dimensionality, and implement the proposed approach using parallelized
routines.

From the simulation studies in Section~\ref{sec:study} we can further draw two
conclusions. First, as reported by other methods such as PUMA, filtering is
important; our EM filtering procedure seems to focus on effectively selecting
true positives at low false positive rates. This feature of our method is
encouraging, since practitioners are often interested in achieving higher
sensitivity by focusing on lower false positive rates. Second, because
we depend on good informative priors to guide the selection of associated
markers, we rely on a judicious choice of hyper-prior parameters, in
particular of the range parameter $\phi$ and how it boosts markers within
neighboring genes that are deemed relevant. It is also important to elicit
gene relevances from well curated databases, e.g.\ MalaCards, and to calibrate
prior strength according to how significant these scores are.

We have shown that our model performs better than most variable selection
methods, but that it can suffer in case of severe model misspecification. As a
way to flag misspecification we suggest to check monotonicity in a measure of
model fit such as PPL as we filter markers using EM.\@ In addition, refining
the EM filtering by using a lower threshold ($< .25$) at each iteration can
help increase performance, especially at lower false positive rates.

When applying the spatial boost model to a real data set, we were able to
confidently isolate at least one marker that has previously been linked to the
trait as well as find another novel interesting marker that may be related to
the trait. This shows that although we can better explore associations jointly
while accounting for gene effects, the spatial boost model still might lack
power to detect associations between diseases and SNPs due to the high
correlation induced by linkage disequilibrium. In the future we plan to
increase the power even further by extending the model to include a data
pre-processing step that attempts to formally correct for the collinearity
between SNPs.

\section*{Acknowledgements}
\label{sec:acknolwedgements}
The authors would like to thank the anonymous reviewers for their valuable
suggestions and constructive comments that helped shape a much improved final
version of the paper. We would also like to thank the editors for their
generous comments and support during the review process. This study makes use
of data generated by the Wellcome Trust Case Control Consortium. A full list
of the investigators who contributed to the generation of the data is
available from \texttt{wtccc.org.uk}. This work was also supported in part by
the ICR-KU International Short-term Exchange Program for Young Researchers,
the JSPS Summer Program 2013 (SP13024), and the NSF EAPSI Program 2013.

\begin{supplement}[id=suppA]
  \sname{Supplement A}
  \stitle{Extended results tables and figures}
  \slink[doi]{COMPLETED BY THE TYPESETTER}
  \sdatatype{.pdf}
  \sdescription{We provide figures and tables to summarize the results of additional simulation studies with less stringent effect sizes as well as the findings in our case study.}
\end{supplement}

\bibliographystyle{chicago}
\bibliography{Spatial_revised_2015}
\end{document}